\pdfoutput=1

\documentclass[twocolumn]{aastex63}
\usepackage{epstopdf}
\usepackage{lineno}
\usepackage{xspace}
\usepackage{graphicx}
\usepackage{amssymb}
\usepackage{amsmath}
\usepackage{natbib}
\usepackage{float}
\usepackage{hyperref}

\shorttitle{{\it Chandra} view on the LINER CGCG\,292$-$057}
\shortauthors{Balasubramaniam et al.}

\begin{document}

\title{{\it Chandra} View of the LINER-type Nucleus in the Radio-Loud Galaxy CGCG\,292$-$057:\\ Ionized Iron Line and Jet--ISM Interactions }

\correspondingauthor{K.~Balasubramaniam}
\email{karthik.balasubramaniam@doctoral.uj.edu.pl}

\author{K.~Balasubramaniam}
\affiliation{Astronomical Observatory of the Jagiellonian University, ul. Orla 171, 30-244 Krak\'ow, Poland}

\author{\L .~Stawarz}
\affiliation{Astronomical Observatory of the Jagiellonian University, ul. Orla 171, 30-244 Krak\'ow, Poland}

\author{V.~Marchenko}
\affiliation{Astronomical Observatory of the Jagiellonian University, ul. Orla 171, 30-244 Krak\'ow, Poland}

\author{M.~Sobolewska},
\affiliation{Harvard Smithsonian Center for Astrophysics, 60 Garden Street, Cambridge, MA 02138, USA}

\author{C.~C.~Cheung}
\affiliation{Naval Research Laboratory, Space Science Division, Washington, DC 20375, USA}

\author{A.~Siemiginowska}
\affiliation{Harvard Smithsonian Center for Astrophysics, 60 Garden Street, Cambridge, MA 02138, USA}

\author{R.~Thimmappa}
\affiliation{Astronomical Observatory of the Jagiellonian University, ul. Orla 171, 30-244 Krak\'ow, Poland}

\author{E.~Kosmaczewski}
\affiliation{Astronomical Observatory of the Jagiellonian University, ul. Orla 171, 30-244 Krak\'ow, Poland}

\begin{abstract}
We present an analysis of the new, deep (94\,ksec) {\it Chandra} ACIS-S observation of radio-loud active galaxy CGCG\,292$-$057, characterized by a LINER-type nucleus and a complex radio structure that indicates intermittent jet activity. On the scale of the host galaxy bulge, we detected excess X-ray emission with a spectrum best fit by a thermal plasma model with a temperature of $\sim 0.8$\,keV. We argue that this excess emission results from compression and heating of the hot diffuse fraction of the interstellar medium displaced by the expanding inner, $\sim 20$\, kpc-scale lobes observed in this restarted radio galaxy. The nuclear X-ray spectrum of the target clearly displays an ionized iron line at $\sim 6.7$\,keV, and is best fitted with a phenomenological model consisting of a power-law (photon index $\simeq 1.8$) continuum absorbed by a relatively large amount of cold matter (hydrogen column density $\simeq 0.7 \times 10^{23}$\,cm$^{-2}$), and partly scattered (fraction $\sim 3\%$) by ionized gas, giving rise to a soft excess component and K$\alpha$ line from iron ions. We demonstrate that the observed X-ray spectrum, particularly the equivalent width of Fe\,\texttt{XXV} K$\alpha$ (of order $0.3$\,keV) can in principle, be explained in a scenario involving a Compton-thin gas located at the scale of the broad-lined region in this source and photoionized by nuclear illumination. We compare the general spectral properties of the CGCG\,292$-$057 nucleus, with those of other nearby LINERs studied in X-rays.
\end{abstract}


\section{Introduction}
\label{sec:intro}

\begin{figure*}[!t]
\centering
\includegraphics[width=\textwidth]{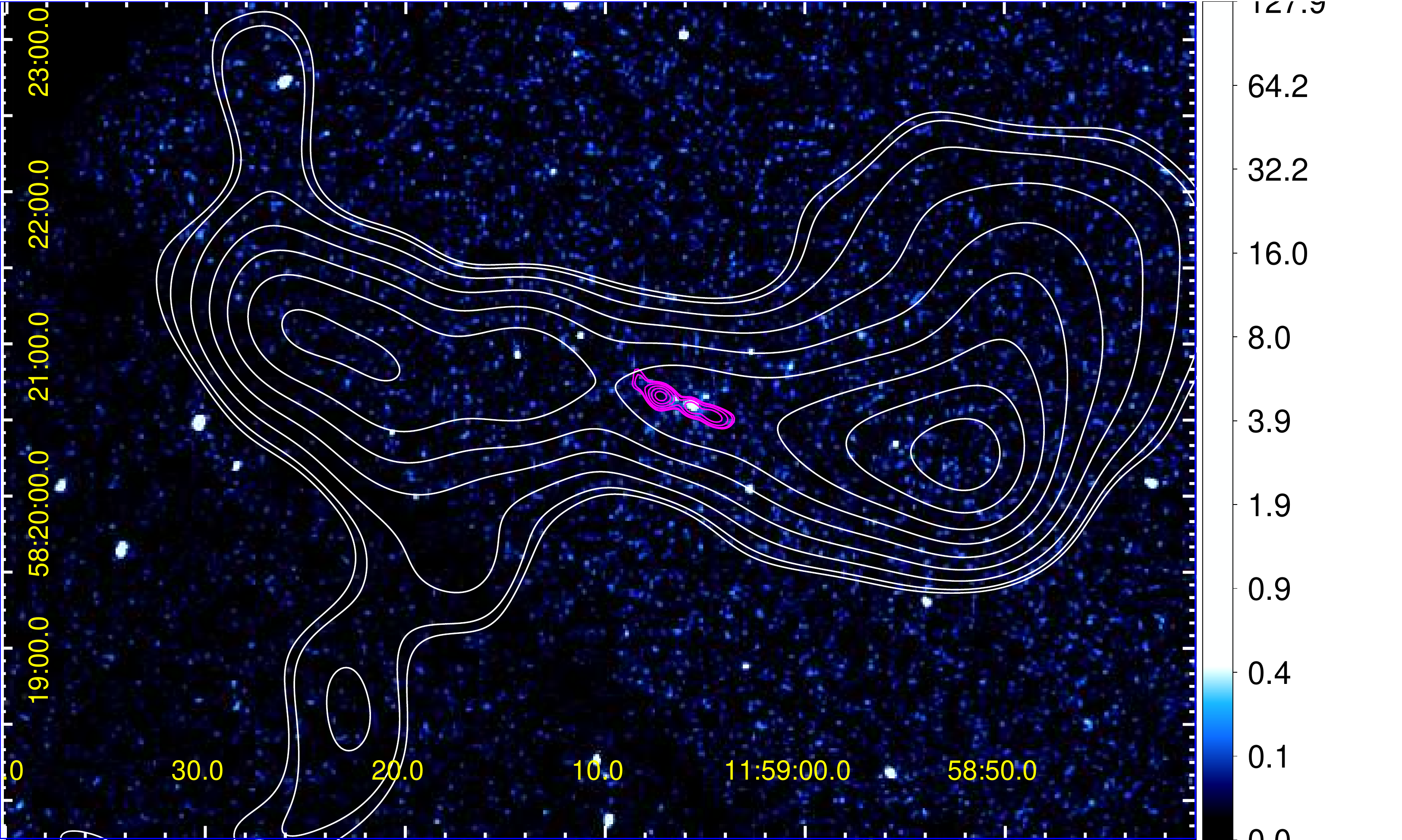}
\caption{The {\it Chandra} ACIS-S image of CGCG\,292$-$057, with the superimposed 606\,MHz radio contours from GMRT. The white contours display the outer structure of the radio source, corresponding to the beam size $45^{\prime\prime} \times 45^{\prime\prime}$, while the magenta contours map the inner radio structure with a beam size $5.9^{\prime\prime} \times 4.6^{\prime\prime}$ with a beam position angle of 46.2\,deg. The contour levels are spaced by a factor of $\sqrt{2}$, starting from 4\,mJy\,beam$^{-1}$ (white) and 1.5\,mJy\,beam$^{-1}$ (magenta). The {\it Chandra} image displays the 0.5--7\,keV counts, smoothed with a 3\,px-radius Gaussian.} 
\label{fig:radio}
\end{figure*}

The nearby galaxy CGCG\,292$-$057 \citep[redshift, $z = 0.054$;][]{Stoughton02} has some unusual properties revealed in studies at radio, infrared, optical, and UV wavelengths by \citet{Koziel12} and \citet{Singh15}. It is a complex, bulge-dominated system viewed nearly face-on, for which the spiral arm tails extend up to $\sim 80$\,kpc, and has a bulge radius of $\sim 30$\,kpc. The concentration index $C \equiv r_{90}/r_{50} =2.81\pm0.03$, which is the ratio of radii containing $90\%$ and $50\%$ of the Petrosian $r$ light, places the galaxy at the border between early-type and late-type galaxies \citep[see][]{Strateva01,Shimasaku01}. Its disturbed optical morphology indicates that it is a relatively recent merger, and the lack of an over-density at the position of the source, as observed in the Tenth Data Release of the Sloan Digital Sky Survey \citep[SDSS;][]{Ahn14}, suggests that it is not associated with any group or cluster of galaxies \citep{Singh15}. Optical spectroscopy of the system reveals 
low-luminosity active nucleus with relatively strong low-ionization emission lines and faint high-ionization lines, consistently with the ``low-ionization nuclear emission-line region'' (LINER) classification \citep[see][for a review]{Ho08}. The mass of the central supermassive black hole (SMBH), estimated based on the mass--stellar velocity dispersion relation \citep{Tremaine02}, or various mass--bulge luminosity relations \citep{Bentz09,McConnell13}, is $\simeq (2.7-5.7)\times 10^8 M_{\odot}$. Based on IR and UV photometry, the star formation rate in the galaxy was estimated as $\simeq (1.5-2.8) \, M_{\odot}$\,yr$^{-1}$ (\citealt{Singh15}, following the scaling relations as given in \citealt{Kennicutt98}).

However, what is particularly interesting about CGCG\,292$-$057 is its multi-scale and multi-component radio morphology, displayed in Figure\,\ref{fig:radio}. On arc-second scales, radio maps of the galaxy reveal a pair of compact lobes extending linearly from the radio core, with a total size of $\sim 20$\,kpc, characteristic of a ``medium symmetric object'' \citep[MSO; see, e.g.,][]{Augusto06}. These lobes are confined to the central bulge and interact directly with the interstellar medium (ISM) of the host. Furthermore, this ``inner'' radio structure is embedded within the arcmin-scale ``outer'' radio lobes, which extend $\sim 250$\,kpc from the galaxy center; the entire radio structure, therefore, may be classified as a ``double-double radio galaxy'' \citep[DDRG; see][]{Mahatma19}. The outer lobes are characterized by a monochromatic 1.4\,GHz radio power of $\sim 2 \times 10^{24}$\,W\,Hz$^{-1}$, which places the source at the low end of the border between the classification of a Fanaroff-Riley type I and type II radio galaxy; these lobes have an additional pair of diffuse wings, giving the system a clear ``X-shaped'' appearance \citep[see][]{Cheung07,Cheung09}. This morphological complexity is most likely reflective of intermittent jet activity of the system's central SMBH. 

\begin{figure*}[!t]
\centering
\includegraphics[width=0.49\textwidth]{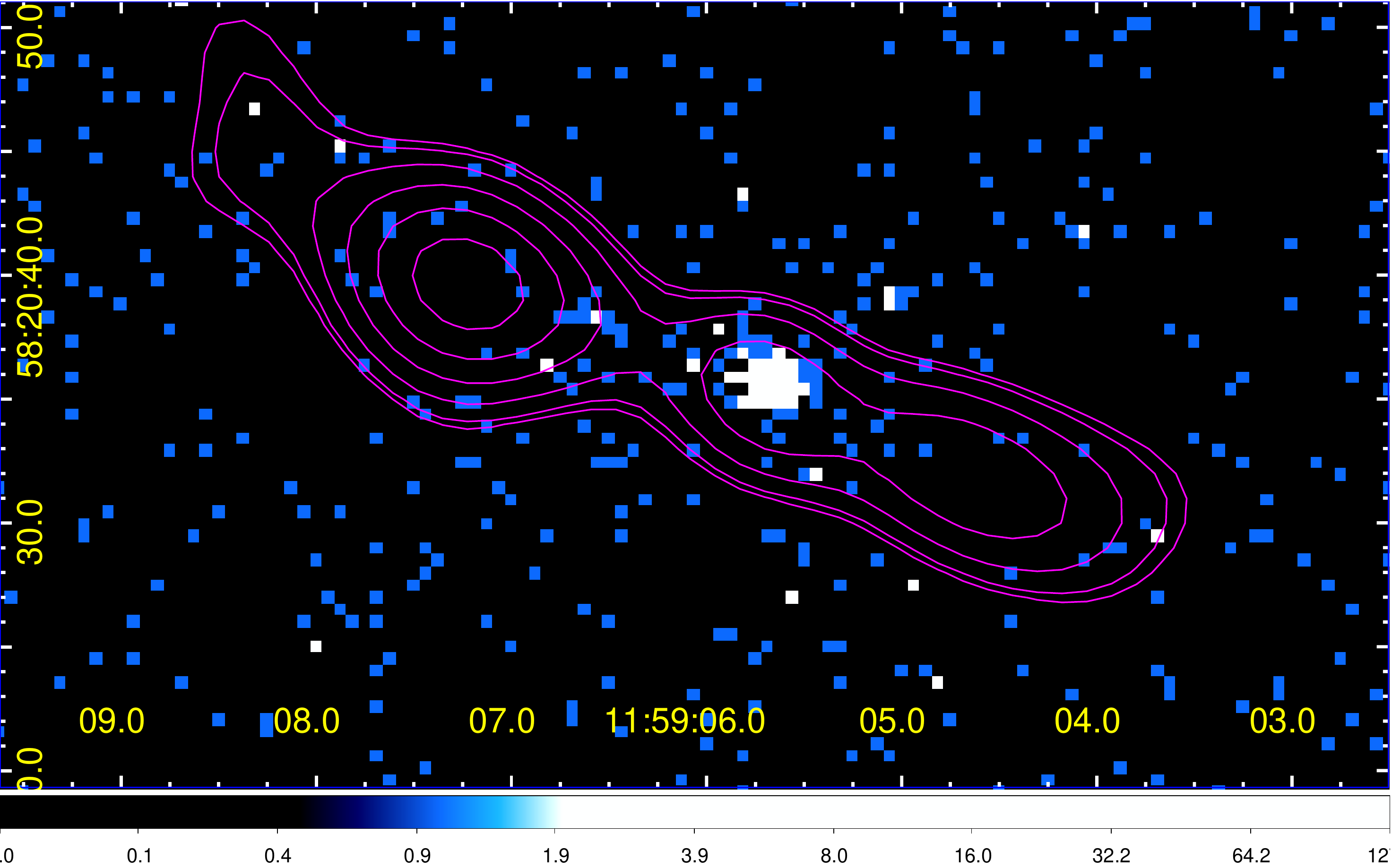}
\includegraphics[width=0.49\textwidth]{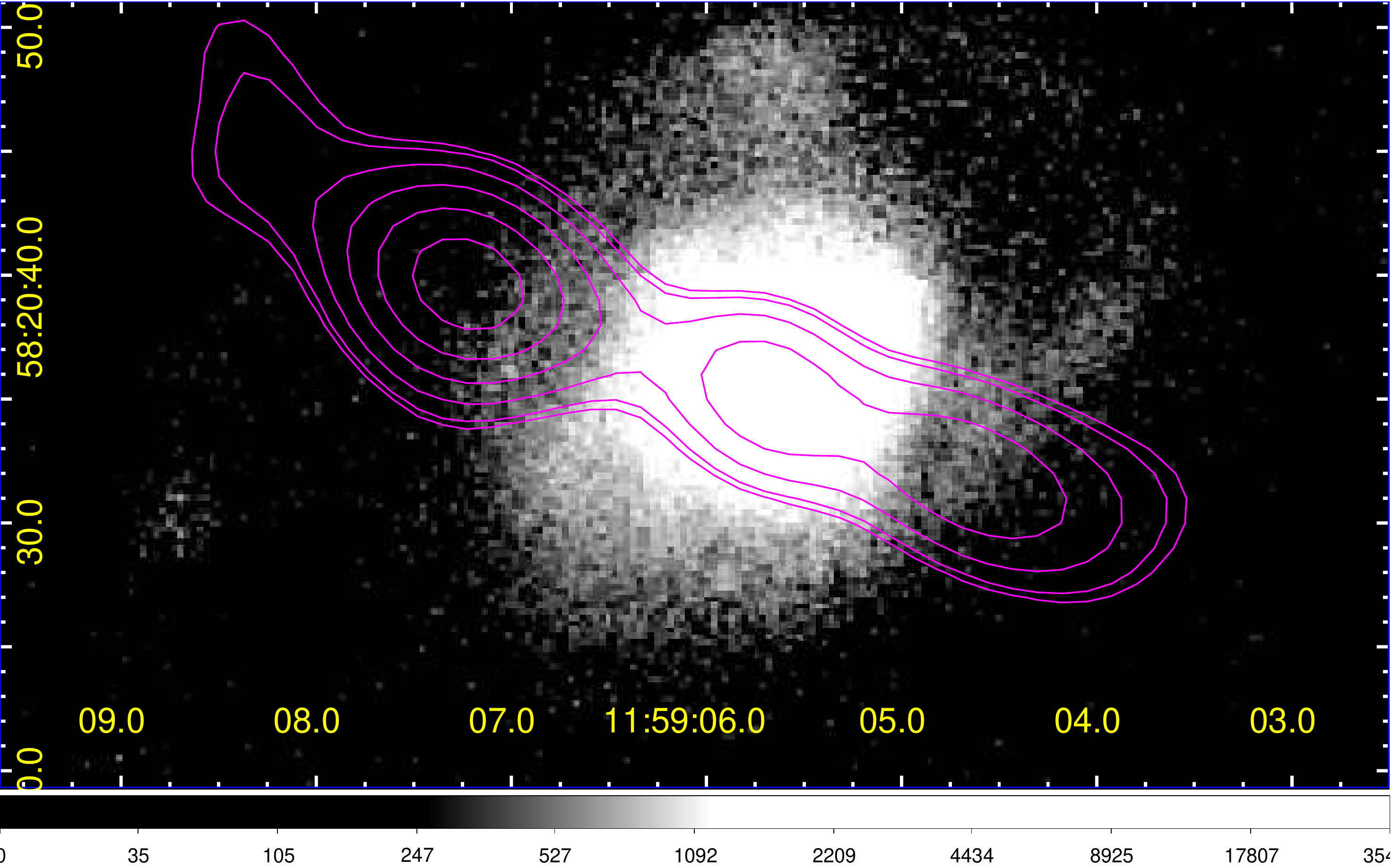}
\caption{The {\it Chandra} ACIS-S image (0.5--7\,keV counts) and the PanSTARRS optical (g-band) image of CGCG\,292$-$057 host galaxy (left and right panels, respectively), both with the superimposed 606\,MHz GMRT contours of the inner radio structure (same as in Figure\,\ref{fig:radio}).}
\label{fig:X}
\end{figure*}

The duty cycle of active galactic nuclei (AGN) is, in general, a subject of ongoing debate. This debate pertains to the various observational manifestations of the activity, timescales, and feedback processes involved. On extended timescales, up to $10^8$\,yrs, major galaxy mergers are broadly considered to be the cause of enhanced accretion flow onto central SMBHs, thus providing material for black hole growth and spin evolution, and fuelling the dominant phase of the central activity \citep[see, e.g.,][]{Hopkins06,Sikora07}. During this phase, as well as in later stages, AGN are likely to pass through subsequent epochs of enhanced and dormant activity. Intermittent behaviour of this type is likely to be caused by minor mergers or instabilities in the accretion flow \citep[e.g.,][]{Czerny09}.

In this framework, the large-scale morphology of radio galaxies with X-shaped radio lobes, could in principle be due to the Lense-Thirring precession, in which the jet axis changes abruptly following the flip in the SMBH spin during an ongoing merging process \citep[see, e.g., the discussion in][and references therein]{Machalski16}. On the other hand, DDRGs, where the inner and outer lobes are well-aligned, are consistent with the development of perturbations in the accretion rate and subsequent intermittent radio activity along the same direction of the outflow \citep[see, e.g.,][and references therein]{Konar19}. Moreover, radio galaxy population studies indicate the existence of far too many compact (linear sizes $\leq 1$\,kpc) and presumably young (ages $< 10^5$\,yrs) sources, as compared to the number of galaxies with old extended radio structures \citep{odea97}. A likely explanation for this inconsistency is that the jet activity is in general highly modulated, and the source is effectively reborn every $\sim 10^4-10^5$\,yrs \citep{Reynolds97}.

Interestingly enough, in the case of CGCG\,292$-$057, one can potentially witness all the aforementioned processes at work, imprinted in the complex morphology of its radio jets/lobes, as well as in its host galaxy. Due to the presence of such peculiar characteristics, we studied this source with the {\it Chandra} X-ray Observatory, making use of the unprecedented combination of the excellent angular resolution and high sensitivity of the ACIS instrument. Below we report on the analysis of the deep {\it Chandra} exposure of the source, including a detailed profile analysis of the surface brightness covering the galactic bulge, as well as the spectroscopy of the unresolved nucleus. Preliminary results of the analysis were presented in \citet{Balasubramaniam17}.

Throughout the paper we assume modern $\Lambda$CDM cosmology with $H_{0}=70$\,km\,s$^{-1}$\,Mpc$^{-1}$, $\Omega _{\rm m}=0.3$, and $\Omega _{\Lambda}=0.7$, so that the luminosity distance of the source for the given redshift $z = 0.054$ is $d_{\rm L} = 240$\,Mpc, and the conversion scale reads as 1.05\,kpc/arcsec.

\section{Chandra Data Analysis} 
\label{sec:chandra}

We observed CGCG\,292$-$057 with {\it Chandra} on 2015 Aug 14 (ObsID 17116; cycle 16) with an exposure totaling 93.8\,ksec. The source (J2000 position, RA\,=\,11h59m5.7s, Decl.\,=\,+58d20$^{\prime}$36$^{\prime\prime}$) was placed at the aim-point of the back-illuminated ACIS-S3 CCD with the data collected in VFAINT mode. The X-ray data analysis was performed with the CIAOv.4.12 software \citep{Fruscione06} using CALDBv.4.9.1. We processed the data by running the CIAO tool \texttt{chandra$\_$repro}. Spectral modeling was performed with Sherpa \citep{Freeman01}, specifically on unbinned spectra, using the Cash and C-stat fitting statistics and the Nelder-Mead or Levenberg-Marquardt optimization methods. 

Below in \S\,\ref{sec:profile} we present the profile analysis of the surface brightness profile for our target, and in \S\,\ref{sec:spectrum} we present spectral modeling of the unresolved nucleus.

\begin{figure*}[!t]
\includegraphics[width=\textwidth]{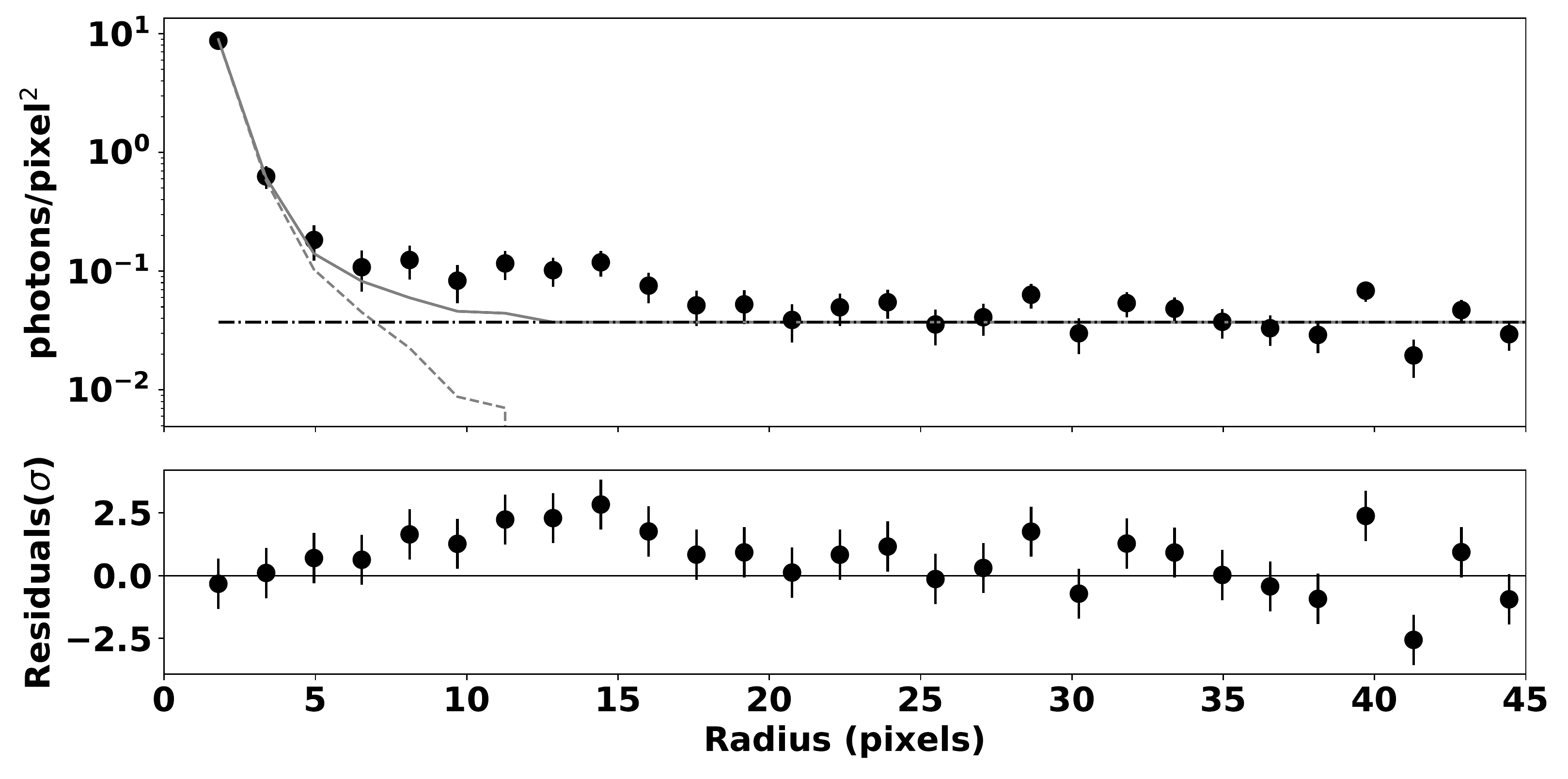}
 \caption{{\it Upper:} X-ray surface brightness profile resulting from extracting the counts of a concentric stack of annular regions centered on the CGCG\,292$-$057 nucleus. The dot-dashed curve denotes the table model for the core PSF, horizontal dotted line corresponds to the constant background, and the solid red curve gives the PSF+background model. {\it Lower:} residuals of the surface brightness profile fitting with the PSF+background model.}
 \label{fig:SB}
\end{figure*} 

\begin{figure}[!th]
\centering
\hbox{\hspace{-0.7cm} \includegraphics[width=1.17\columnwidth]{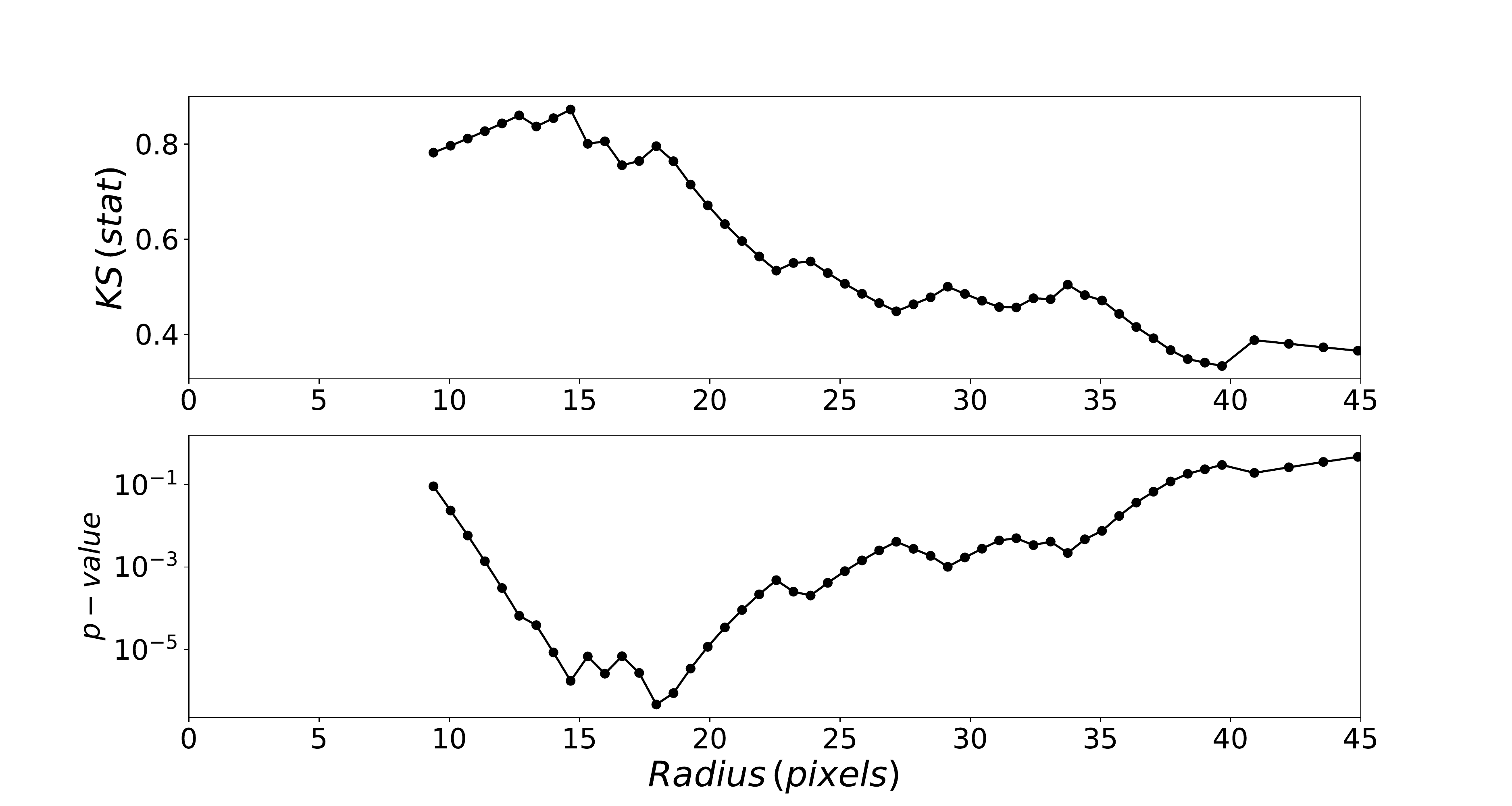}}
 \caption{KS statistics and $p$-values corresponding to the null hypothesis that the samples of surface brightness profiles within the annuli 10\,px---X\,px and X\,px---45\,px were drawn from the same distribution, as functions of a running variable X\,$\in$\,(10\,px, 45\,px).}
\label{fig:KS}
\end{figure}

\subsection{Surface Brightness Profile}
\label{sec:profile}

The target was detected with a total number of 393 net counts within the $0.5-7.0$\,keV range and 5\,px source extraction radius. Figure\,\ref{fig:X} displays the ACIS-S image of the central region of CGCG\,292$-$057, along with overlaid 606\,MHz contours from the Giant Metrewave Radio Telescope (GMRT) marking the inner radio structure of the source \citep[see][]{Koziel12}. As shown, the active nucleus of the system is clearly detected in X-ray, surrounded by diffuse emission with a low-surface brightness.

We extracted the radial profile of the counts from the exposure-corrected {\it Chandra} map using concentric annular regions centered on the CGCG\,292$-$057 nucleus. The resultant surface brightness profile was fit with a model including the core PSF and a constant background, with both the PSF normalization and the background (constant) amplitude set free. Instead of a 2D Gaussian approximation for the core PSF, we built a table model from the averaging of over 100 PSF simulations, as described in Appendix\,\ref{appendix}. The corresponding best-fit model along with the residuals are given in Figure\,\ref{fig:SB}. As shown, positive residuals are seen within a range starting at around $10$\,px extending to $\lesssim 20$\,px from the core, signalling the presence of an additional, extended, X-ray emission component. 

As the significance of the excess may not seem high, we calculated the Kolmogorov-Smirnov (KS) test to quantify the probability that the samples of surface brightness profiles within the annuli 10\,px---X\,px, and X\,px---45\,px, with a running variable X\,$\in$\,(10\,px, 45\,px), were drawn from the same distribution. The results of the test, i.e. the KS statistics and the corresponding $p$-values as functions of X, are presented in Figure\,\ref{fig:KS}. As shown, $p< 10^{-6}$ for X$\,\simeq 18$\,px, implying that the emission seen between 10\,px and 18\,px --- which is beyond the $3\sigma$ radius for the photon counts from the unresolved core (see Appendix\,\ref{appendix}) --- is significantly distinct from the constant background present outside of the 18\,px radius. 

Notably, this excess emission, located at physical distances of $\sim 5$\,kpc from the core (up to $\sim 9$\,kpc, for the conversion scale $\sim 1$\,kpc/$^{\prime\prime}$ and 1\,px\,$\simeq 0.5^{\prime\prime}$), coincides with the edges of the inner radio structure (MSO) of CGCG\,292$-$057 (the radio lobe radius $\sim 10$\,kpc). It is, therefore, natural to relate this feature with the compact radio jets/lobes in the system. In particular, the X-ray excess could either be non-thermal emission of the compact lobes \citep{Stawarz08}, or a hot diffuse phase of the ISM, compressed and heated by the expanding young radio lobes \citep{Reynolds01}.

In order to investigate possible differences in the position of the excess emission along and across the major axis of the inner radio lobes, we have also extracted the radial profiles of the net counts separately for the four quadrants of the concentric annular regions, oriented in such a way that the East and West quadrants correspond to the surface brightness profiles \emph{along} the jet axis, while the North and South quadrants to the surface brightness profiles \emph{across} the jet axis. The results are presented in Appendix\,\ref{appendixB}. As shown, due to the limited photon statistics in separate quadrants, even with the size of the annulus bins increased from 2\,px up to 7\,px, the differences are not significant statistically, although there is a tentative evidence of an elongation of the X-ray emission along the jet direction (in particular, in the East quadrant when compared to the South and North ones).

\begin{deluxetable*}{llcc}[t!]
\tabletypesize{\footnotesize}
\tablecaption{Results of the simultaneous fitting of the ISM ($=$\,source) and the BKG ($=$\,background) regions. \label{tab:ISM}}
\tablewidth{0pt}
\tablehead{\colhead{Region} & \colhead{Model$^{\dagger}$/Parameter} & \colhead{Value/Value with $1\,\sigma$ errors} & \colhead{Unit$^{\ddagger}$} }
\startdata
{\bf BKG:} &\multicolumn{3}{l}{\underline{\texttt{powerlaw}}} \\
 & Photon index $\Gamma_{\rm bcg}$ & $0.87 \pm 0.15$ &   \\
 & PL amplitude & $(8.62 \pm 1.48) \times 10^{-7}$ & keV$^{-1}$\,cm$^{-2}$\,s$^{-1}$ at 1\,keV  \\
{\bf ISM:} & \multicolumn{3}{l}{\underline{\texttt{zphabs * xsapec}}}  \\
 & Hydrogen column density $N_{\rm H}^{\rm (ISM)}$ & $0.53^{+0.14}_{-0.14}$ & $10^{22}$\,cm$^{-2}$  \\
 & Temperature $kT_{\rm ISM}$ & $0.81^{+0.23}_{-0.35}$ & keV  \\
 & Normalization & $(2.21\pm 0.65)\times 10^{-6}$ & $10^{-14} (1+z)^2 n_e n_{\rm H} V/4\pi d_{\rm L}^2$  \\
& {\bf C-stat./DOF} & $455/887$ & \\
\hline
{\bf BKG:} & \multicolumn{3}{l}{\underline{\texttt{powerlaw}}} \\
 & Photon index $\Gamma_{\rm bcg}$ & $0.87 \pm 0.18$ &   \\
 & PL amplitude & $(8.44 \pm 1.64) \times 10^{-7}$ & keV$^{-1}$\,cm$^{-2}$\,s$^{-1}$ at 1\,keV  \\
{\bf ISM:} & \multicolumn{3}{l}{\underline{\texttt{zphabs * powerlaw}}} \\
 & Hydrogen column density $N_{\rm H}^{\rm (ISM)}$ & $0.35^{+0.35}_{-} \times 10^{-7}$ & $10^{22}$\,cm$^{-2}$  \\
 & Photon index $\Gamma_{\rm ISM}$ & $4.01 \pm 1.78$ &   \\
 & PL amplitude & $(1.87 \pm 1.86) \times 10^{-7}$ & keV$^{-1}$\,cm$^{-2}$\,s$^{-1}$ at 1\,keV  \\
& {\bf C-stat./DOF} & $455/887$ &
\enddata
\tablecomments{$^{\dagger}$ all the models include the Galactic hydrogen column density $N_{\rm H,\,Gal} = 1.43 \times 10^{20}$\,cm$^{-2}$; thermal model assumes solar abundance; $^{\ddagger}$ normalization for the model \texttt{xsapec} assumes uniform ionized plasma with electron and H number densities $n_e$ and $n_{\rm H}$, respectively, and volume $V$ in cgs units.}
\end{deluxetable*}

\begin{deluxetable*}{llcc}[!th]
\tabletypesize{\footnotesize}
\tablecaption{Results of the simultaneous fitting of the AGN ($=$\,source) and the BKG ($=$\,background) regions. \label{tab:AGN}}
\tablewidth{0pt}
\tablehead{\colhead{Model$^{\star}$} & \colhead{Model parameter} & \colhead{Value/Value with $1\,\sigma$ errors} & \colhead{Units$^{\diamond}$}}
\startdata
 \multicolumn{4}{l}{{\bf Model A:} \quad \underline{\texttt{zphabs * powerlaw}}} \\
{\bf C-stat./DOF:} 817/887 & Hydrogen column density $N_{\rm H}$ & $0.028^{\dagger}$ & $10^{22}$\,cm$^{-2}$  \\
 & Photon index $\Gamma$ & $-0.37 \pm0.04$  &   \\
 & PL amplitude & $(1.28 \pm 0.06) \times 10^{-6}$  & keV$^{-1}$\,cm$^{-2}$\,s$^{-1}$ at 1\,keV  \\
 & BKG Photon index $\Gamma_{\rm bcg}$ & $0.86 \pm0.17$  &    \\
 & BKG PL amplitude & $(8.23 \pm 1.41) \times 10^{-7}$ & keV$^{-1}$\,cm$^{-2}$\,s$^{-1}$ at 1\,keV \\
\hline
\multicolumn{4}{l}{{\bf Model B:} \quad  \underline{\texttt{zphabs * powerlaw + xszgauss}}} \\
{\bf C-stat./DOF:} 815/884 & Hydrogen column density $N_{\rm H}$ & $0.052^{\dagger}$ & $10^{22}$\,cm$^{-2}$  \\
 & Photon index $\Gamma$ & $-0.31 \pm 0.05$ &   \\
 & PL amplitude & $(1.35 \pm 0.08) \times 10^{-6}$  & keV$^{-1}$\,cm$^{-2}$\,s$^{-1}$ at 1\,keV  \\
  & Source frame line energy $E_l$ & $6.76^{\dagger}$ & keV \\
  & Line normalization & ${3.17 \times 10^{-7}}^{\dagger}$ & cm$^{-2}$\,s$^{-1}$ \\
& BKG Photon index $\Gamma_{\rm bcg}$ & $0.86 \pm 0.15$ &   \\
 & BKG PL amplitude & $(8.23 \pm 1.23) \times 10^{-7}$ & keV$^{-1}$\,cm$^{-2}$\,s$^{-1}$ at 1\,keV  \\
\hline
\multicolumn{4}{l}{{\bf Model C:}  \quad \underline{\texttt{zphabs * xsapec}}} \\
{\bf C-stat./DOF:} 867/884 & Hydrogen column density $N_{\rm H}$ & $3.15 \pm 0.36$  & $10^{22}$\,cm$^{-2}$ \\
 & Temperature $kT$ & $ >10^{\dagger}$ & keV \\
 & Normalization & $ {1.05 \times 10^{-4}}^{\dagger}$ & $10^{-14} (1+z)^2 n_e n_{\rm H} V/4\pi d_{\rm L}^2$  \\
 & BKG Photon index $\Gamma_{\rm bcg}$ &$1.08 \pm 0.19$  &   \\
 & BKG PL amplitude & $(10.65 \pm 1.92) \times 10^{-7}$  & keV$^{-1}$\,cm$^{-2}$\,s$^{-1}$ at 1\,keV  \\
\hline
\multicolumn{4}{l}{{\bf Model D:}  \quad \underline{\texttt{zphabs(1) * xsapec + zphabs(2) * powerlaw}}} \\
{\bf C-stat./DOF:} 784/884 & Hydrogen column density $N_{\rm H}^{(1)}$ & $1.38 \pm 0.39$ & $10^{22}$\,cm$^{-2}$ \\
 & Temperature $kT$ & $0.12 \pm 0.06$ & keV  \\
 & Normalization & $0.00664^{+0.025}_{-} $ & $10^{-14} (1+z)^2 n_e n_{\rm H} V/4\pi d_{\rm L}^2$  \\
  & Hydrogen column density $N_{\rm H}^{(2)}$ & $3.96 \pm 0.96$ & $10^{22}$\,cm$^{-2}$ \\
 & Photon index $\Gamma$ & $1.09 \pm 0.35$ &   \\
 & PL amplitude & $(1.45 \pm 0.82) \times 10^{-6}$  & keV$^{-1}$\,cm$^{-2}$\,s$^{-1}$ at 1\,keV \\
 & BKG Photon index $\Gamma_{\rm bcg}$ & $0.86 \pm 0.19$  &   \\
 & BKG PL amplitude & $(8.24 \pm 1.61) \times 10^{-6}$ & keV$^{-1}$\,cm$^{-2}$\,s$^{-1}$ at 1\,keV  \\
\hline
\multicolumn{4}{l}{{\bf Model E:}  \quad \underline{\texttt{zphabs(1) * xsapec(1) + zphabs(2) * xsapec(2)}}} \\
{\bf C-stat./DOF:} 785/884 & Hydrogen column density $N_{\rm H}^{(1)}$ & $5.01 \pm 0.43$ & $10^{22}$\,cm$^{-2}$ \\
 & Temperature $kT^{(1)}$ & $17.27 \pm 3.37$ & keV  \\
 & Normalization$^{(1)}$ & $(1.13 \pm 0.08)\times 10^{-4}$ & $10^{-14} (1+z)^2 n_e n_{\rm H} V/4\pi d_{\rm L}^2$  \\
 & Hydrogen column density $N_{\rm H}^{(2)}$ & $1.09 \pm 0.06$ & $10^{22}$\,cm$^{-2}$ \\
 & Temperature $kT^{(2)}$ & $0.22 \pm 0.01$ & keV  \\
 & Normalization$^{(2)}$ & ${1.5 \times 10^{-4}}^{\dagger}$  & $10^{-14} (1+z)^2 n_e n_{\rm H} V/4\pi d_{\rm L}^2$ \\
 & BKG Photon index $\Gamma_{\rm bcg}$ & $0.86 \pm 0.17$ &   \\
 & BKG PL amplitude & $(8.22 \pm 1.39) \times 10^{-7}$ & keV$^{-1}$\,cm$^{-2}$\,s$^{-1}$ at 1\,keV  \\
\hline
\multicolumn{4}{l}{{\bf Model F:}  \quad \underline{\texttt{zphabs(1) * (1-fsc) * powerlaw + zphabs(2) * (fsc * powerlaw +  xszgauss)}}}\\
{\bf C-stat./DOF:} 784/882  & Hydrogen column density $N_{\rm H}^{(1)}$ & $6.86_{-0.58}^{+0.67}$ & $10^{22}$\,cm$^{-2}$  \\
 & Scattering fraction $f_{sc}$ & $0.033_{-0.009}^{+0.011}$ & \\
 & Hydrogen column density $N_{\rm H}^{(2)}$ & $0.09_{-}^{+0.13}$ & $10^{22}$\,cm$^{-2}$ \\
 & Photon index $\Gamma$ & $1.80_{-0.05}^{+0.42}$  &  \\
 & PL amplitude & $(4.57_{-2.16}^{+4.54})\times 10^{-5}$ & keV$^{-1}$\,cm$^{-2}$\,s$^{-1}$ at 1\,keV  \\
  & Source frame line energy $E_l$ &  $6.76_{-0.03}^{+0.03}$ & keV \\
   & Line normalization & $(5.43_{-2.32}^{+2.81})\times 10^{-7}$ & cm$^{-2}$\,s$^{-1}$ \\
  & Equivalent width of the line EW$^{\ddagger}$ & $330^{+210}_{-160}$ & eV \\
 & BKG Photon index $\Gamma_{\rm bcg}$ & $0.86_{-0.12}^{+0.14}$ &   \\
 & BKG PL amplitude & $(8.25_{-1.51}^{+1.71})\times 10^{-7}$ & keV$^{-1}$\,cm$^{-2}$\,s$^{-1}$ at 1\,keV 
\enddata
\tablecomments{$^{\star}$ all the models include the Galactic hydrogen column density $N_{\rm H,\,Gal} = 1.43 \times 10^{20}$\,cm$^{-2}$; thermal models assume solar abundance; in the models including redshifted Gaussian line the source-frame line width frozen at $\sigma_l = 10$\,eV; \\ $^{\diamond}$ normalization for the model \texttt{xsapec} assumes uniform ionized plasma with electron and H number densities $n_e$ and $n_{\rm H}$, respectively, and volume $V$, both in cgs units;\\
$^{\dagger}$ lower and upper $1\,\sigma$ bounds unconstrained when using Nelder-Mead optimization method;\\
$^{\ddagger}$ EW calculated against the direct PL continuum; $1\sigma$ bounds calculated with \texttt{error=True} option in Sherpa, using $10^4$ number of draws.}
\end{deluxetable*}

\begin{figure*}[!th]
\centering
\includegraphics[width=0.495\textwidth]{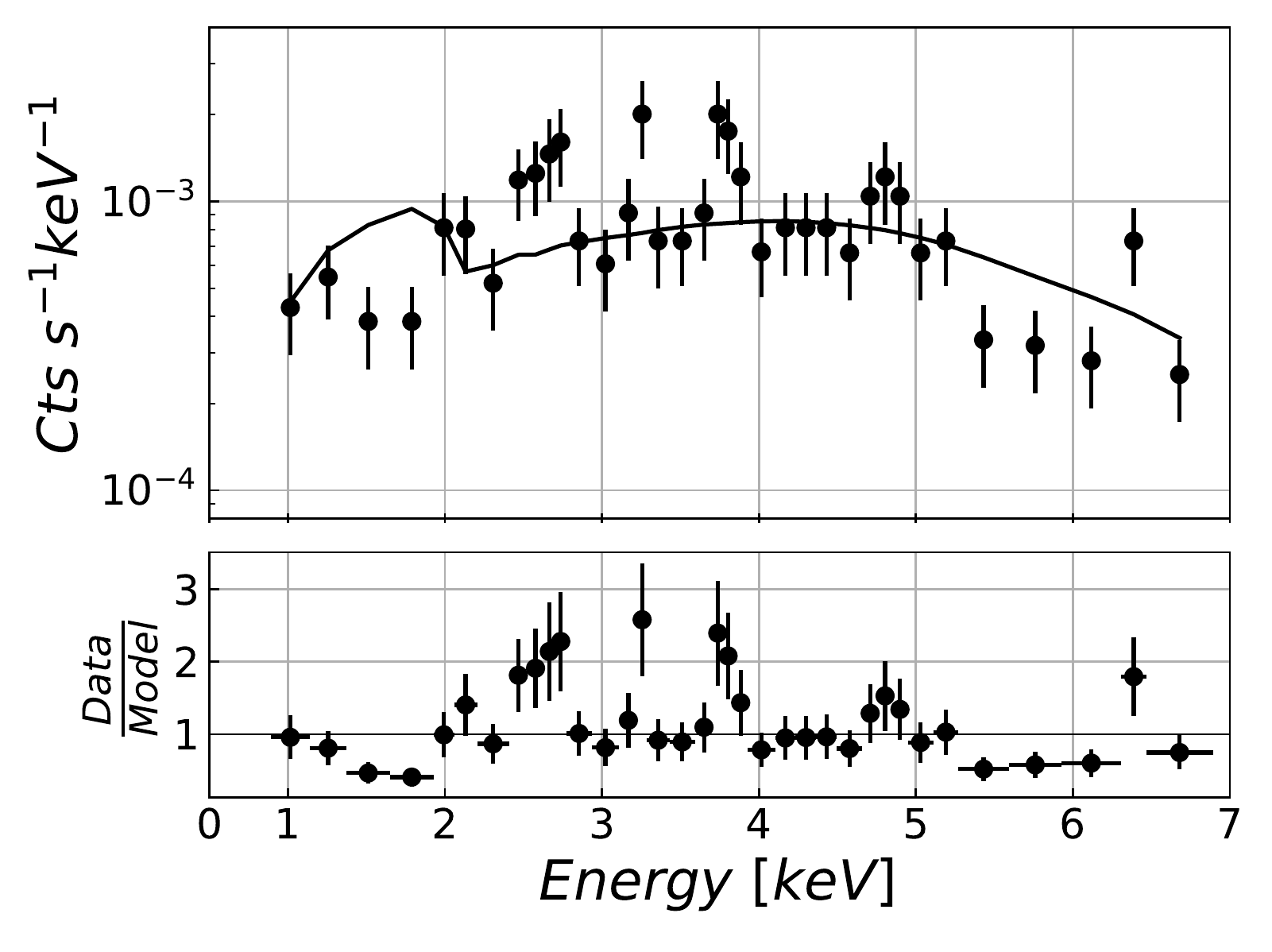}
\includegraphics[width=0.495\textwidth]{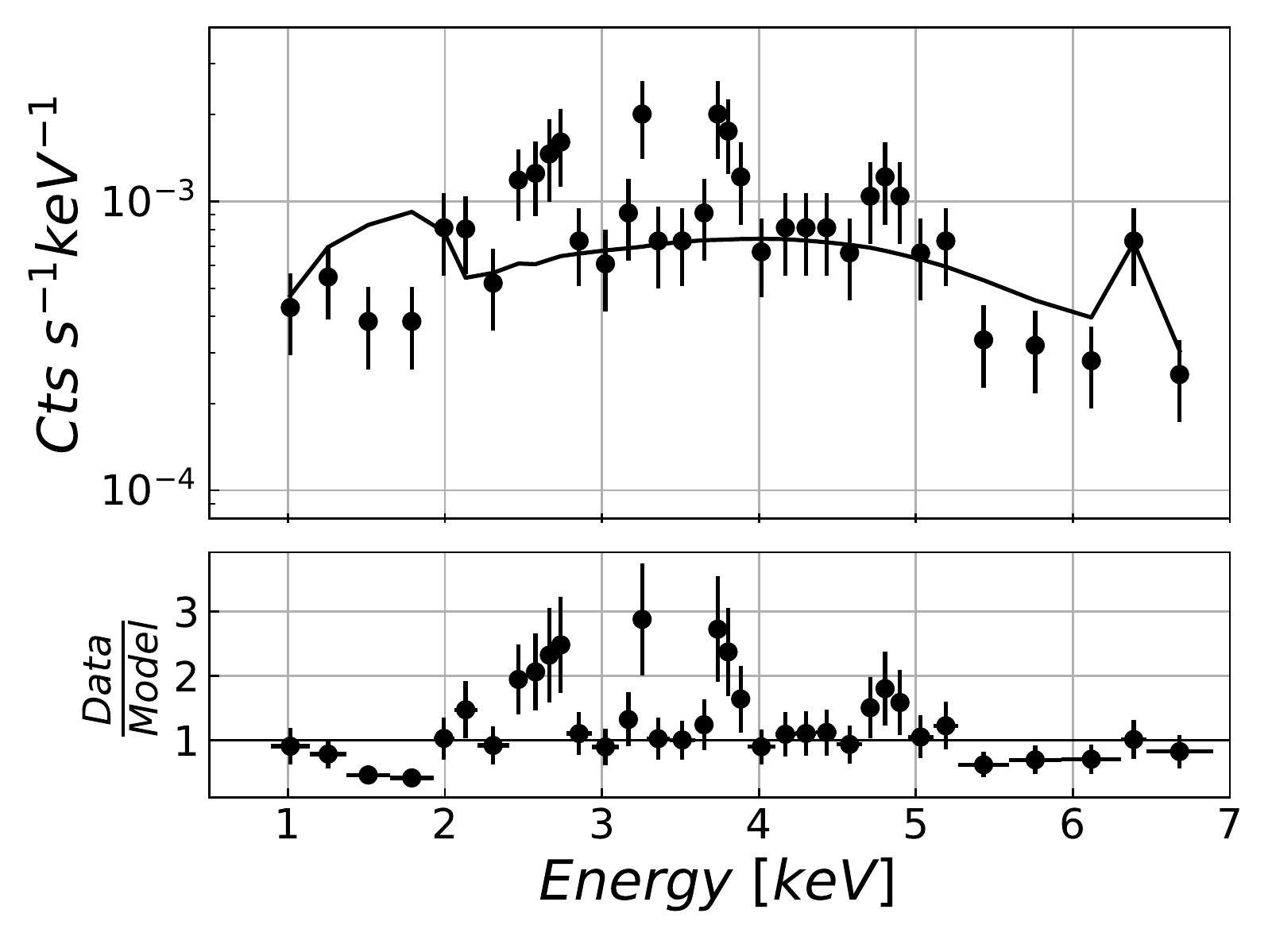}
\includegraphics[width=0.495\textwidth]{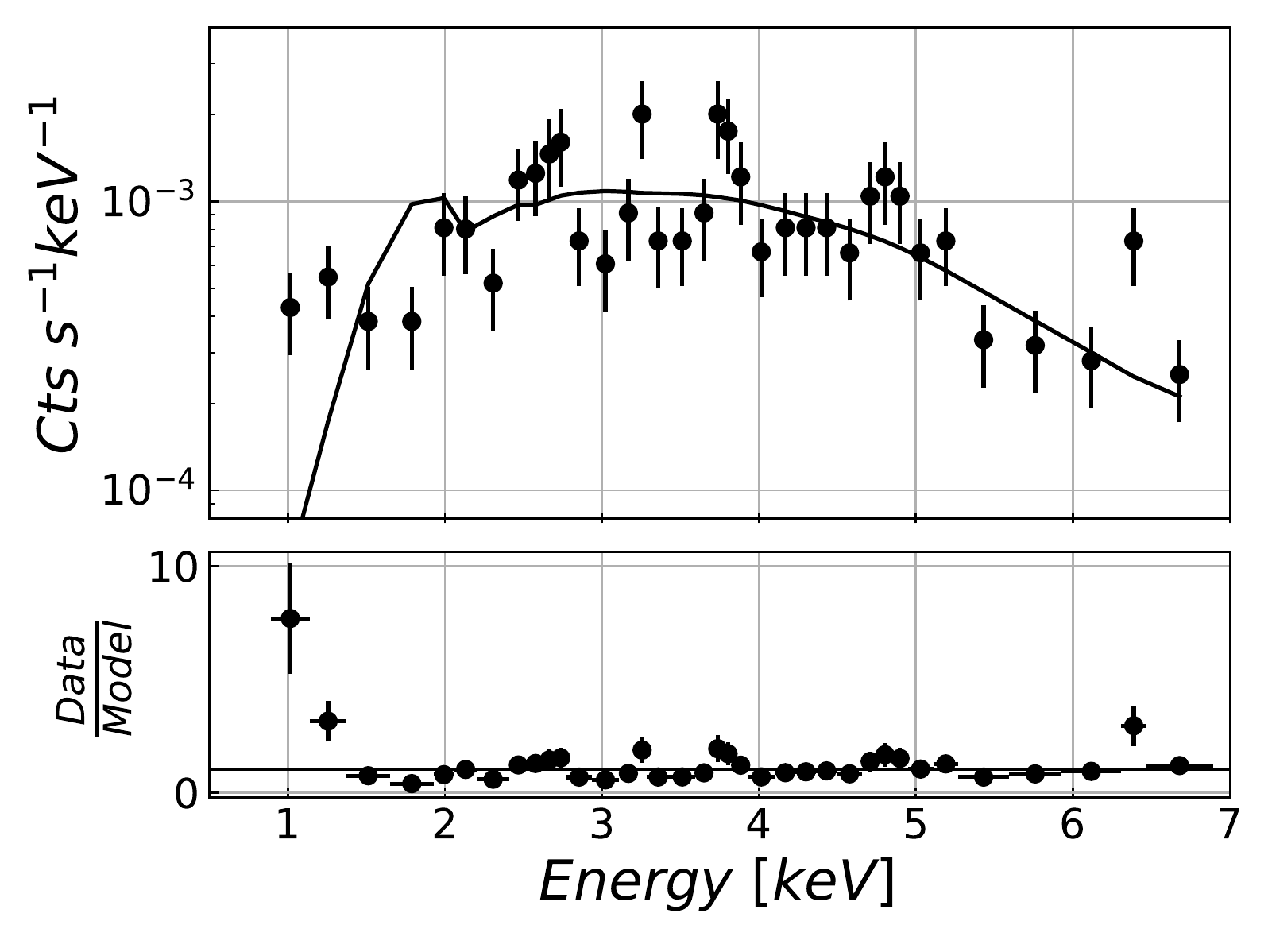}
\includegraphics[width=0.495\textwidth]{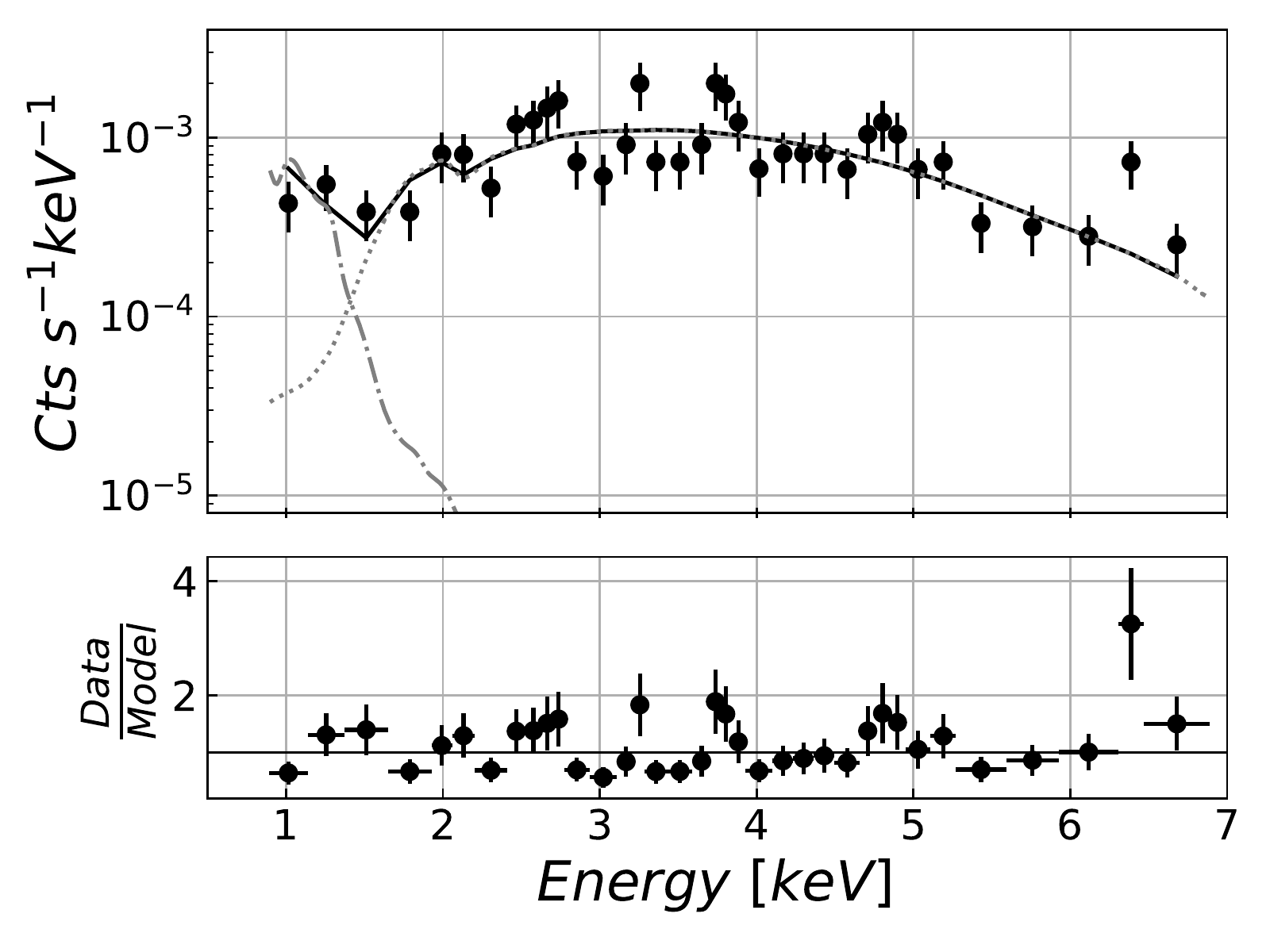}
\includegraphics[width=0.495\textwidth]{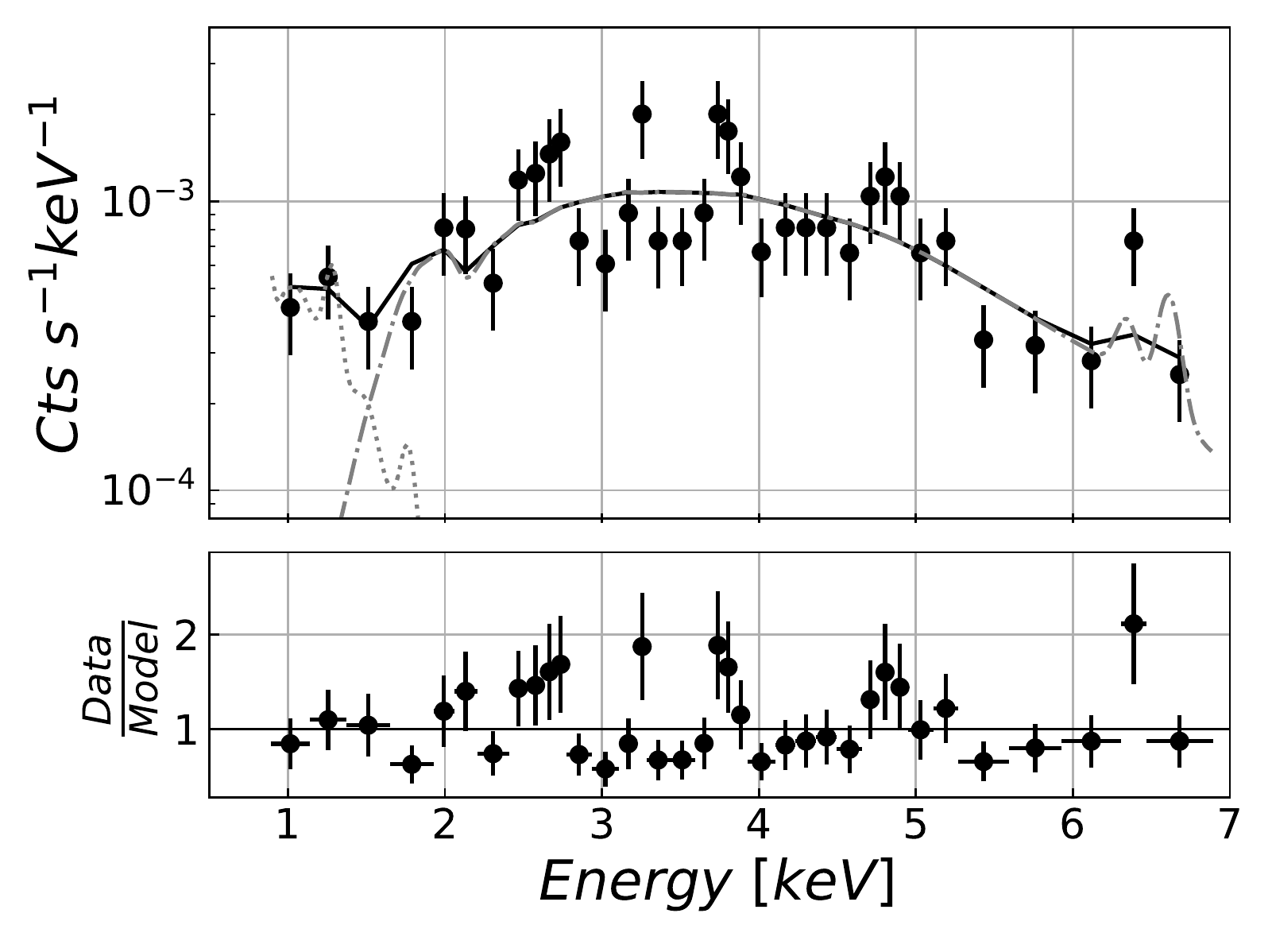}
\includegraphics[width=0.495\textwidth]{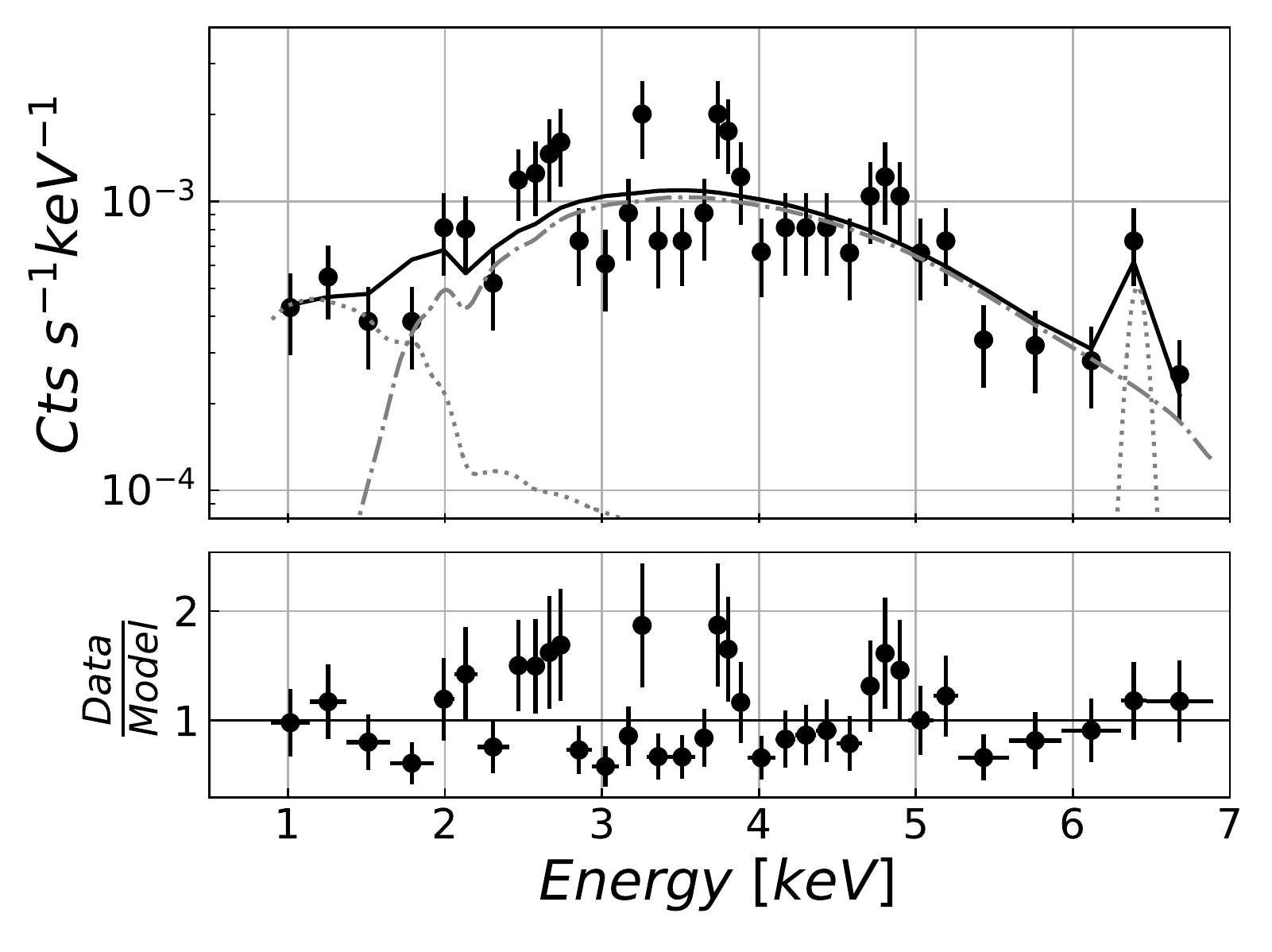}
\vspace{0.1cm} 
\caption{The $0.5-7$\,keV spectrum of the AGN emission component within CGCG\,292$-$057 (binned to reach a signal-to-noise ratio of 3 in each bin, for representation purposes only, as the fits were performed on the unbinned spectra), fitted with various emission modelsl: A (top left panel) a power-law model ``PL'', B (top right) a power-law with a Gaussian line ``PL+G'', C (middle left) a thermal model ``APEC'', D (middle right) a combination of thermal and power-law components ``APEC+PL'', E (bottom left) a combination of two thermal components ``2\,$\times$APEC'', and F (bottom right) a scattered power-law with a Gaussian line ``Scattered PL+G''; see Table\,\ref{tab:AGN} for the resulting best-fit model parameters.}
 \label{fig:models}
\end{figure*}

\subsection{Spectral Analysis}
\label{sec:spectrum}

For spectral modeling, we defined the following extraction regions corresponding to the emission components identified in the X-ray surface brightness profile in CGCG\,292$-$057 from the previous section (\ref{sec:profile}; cf., Figure\,\ref{fig:SB}).
\begin{itemize}
\item[(i)] The ``AGN'' region is extracted using a circular region with a 5\,px radius centered on the central X-ray peak. The region includes at least $95.45\%$ of photons emitted by the active nucleus (for all the simulated PSFs; see Appendix\,\ref{appendix} and in particular the $2\sigma$ confidence line in Figure\,\ref{fig:PSF}), and on the other hand minimizes contamination from the extended (excess) emission component.
\item[(ii)] The ``ISM'' emission is defined with an annulus with an inner radius of 10\,px and an outer radius of 15\,px. This annulus contains the bulk of the excess extended emission component with only minimal pollution from the AGN ($<0.27\%$ of the photons from the core; see the $3\sigma$ line in Figure\,\ref{fig:PSF}).
\item[(iii)] The background, ``BKG'', was taken from an annulus with the inner radius of 20\,px and the outer radius of 30\,px, and corresponding to the outer regions characterized by a flat (constant) X-ray surface brightness. 
\end{itemize}
The spectra for these regions and the corresponding calibration files (ARF and RMF) were extracted with the CIAO script \texttt{specextract}.

First, we simultaneously fitted the $0.5-7.0$\,keV spectra of the ISM region ($=$\,source) and the BKG region ($=$\,background), assuming an absorbed thermal model for the former (\texttt{zphabs * xsapec}, redshift $z=0.054$) and an unabsorbed power-law model for the latter, both  including the Galactic hydrogen column density $N_{\rm H,\,Gal} = 1.43 \times 10^{20}$\,cm$^{-2}$. In the thermal model, we have also fixed the plasma abundance to solar. The results of the fitting (including $1\,\sigma$ errors calculated using the Nelder-Mead optimization method) are summarized in Table\,\ref{tab:ISM}. As follows, the excess emission seen in the ISM region, could be well represented by a thermal emission of a hot gas with the temperature of $\sim 0.8$\,keV and non-negligible intrinsic absorption $\sim 5 \times 10^{21}$\,cm$^{-2}$.

Next we repeated the fit, replacing the absorbed thermal model for the ISM region with an absorbed (\texttt{zphabs}, $z=0.054$) power-law model. The results of this fitting are again summarized in Table\,\ref{tab:ISM}. In this case, the intrinsic absorption turns out to be negligible, and the poorly constrained photon index reads as relatively steep, $\sim 4.0 \pm 1.8$. Consequently, we conclude that the thermal fit to the ISM region (corresponding to the excess in the surface brightness profile over the constant background), returns more plausible parameters, and as such is favored.

Finally, we performed simultaneous fitting for the $0.5-7.0$\,keV spectra of the AGN ($=$\,source) and the BKG ($=$\,background) regions, assuming various emission models for the former (at $z=0.054$), and, as before, an unabsorbed power-law model for the latter, all including the Galactic hydrogen column density $N_{\rm H,\,Gal} = 1.43 \times 10^{20}$\,cm$^{-2}$. The emission models assumed for the AGN include: (A) an absorbed power-law, (B) an absorbed power-law with a Gaussian line, (C) an absorbed thermal model, (D) a combination of an absorbed thermal model with an absorbed power-law, with different absorbers, (E) a combination of two absorbed thermal models, again allowing for two different absorbers, and (F) an absorbed scattered power-law model with a Gaussian line and two different absorbers. In all the thermal models, we froze the abundance at the solar value; in the models including a redshifted Gaussian line, the source-frame line width was frozen at $\sigma_l = 10$\,eV. The results of the fitting (including $1\,\sigma$ errors calculated using the Nelder-Mead optimization method) are summarized in Table\,\ref{tab:AGN} and Figure\,\ref{fig:models}. 

Note that clear residuals, reminiscent of line emission, are visible in the AGN spectrum below 5\,keV; if real, those could signal, e.g., the presence of a collisionally ionized gas within the central parts of the CGCG\,292$-$057 host \citep[see in this context][]{Beuchert18}. Detailed modeling of those features is, however, beyond the scope of the current paper, since such an analysis and line identification are all hampered by an insufficient energy resolution and limited photon statistics of the currently available {\it Chandra} dataset.

\begin{figure*}[!t]
\centering
\hbox{\hspace{-0.6cm}\includegraphics[width=0.6\textwidth]{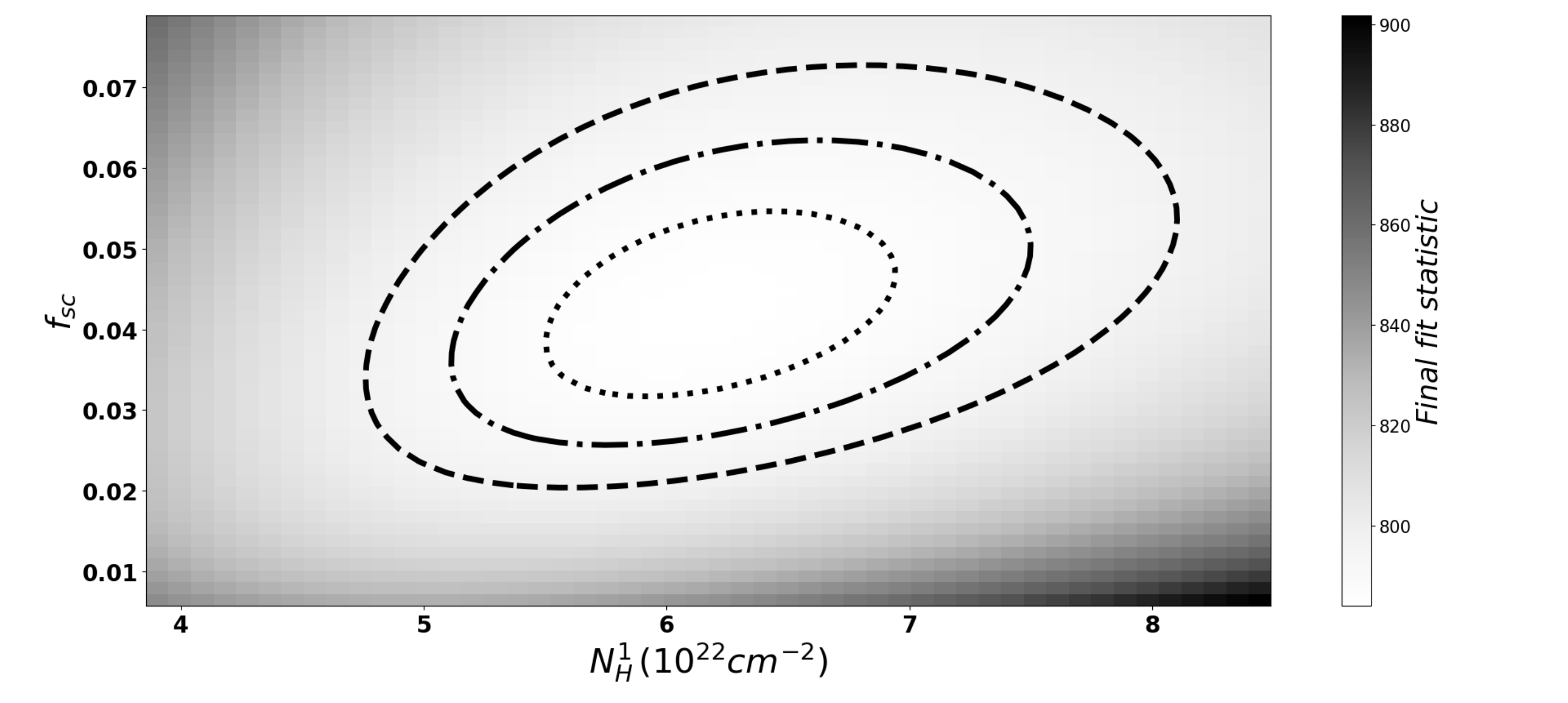}
\hspace{-1.9cm} \includegraphics[width=0.6\textwidth]{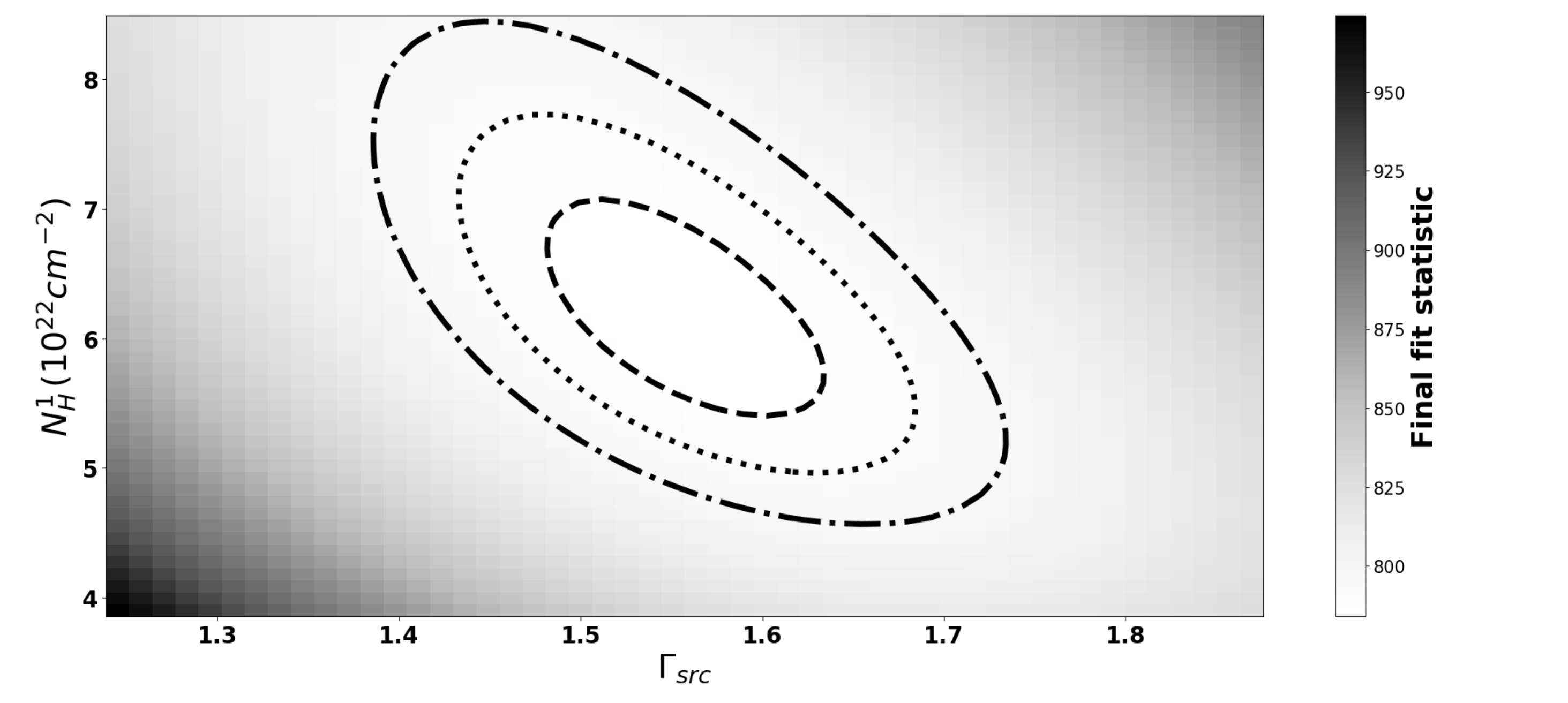}}
\hbox{\hspace{-0.6cm}\includegraphics[width=0.6\textwidth]{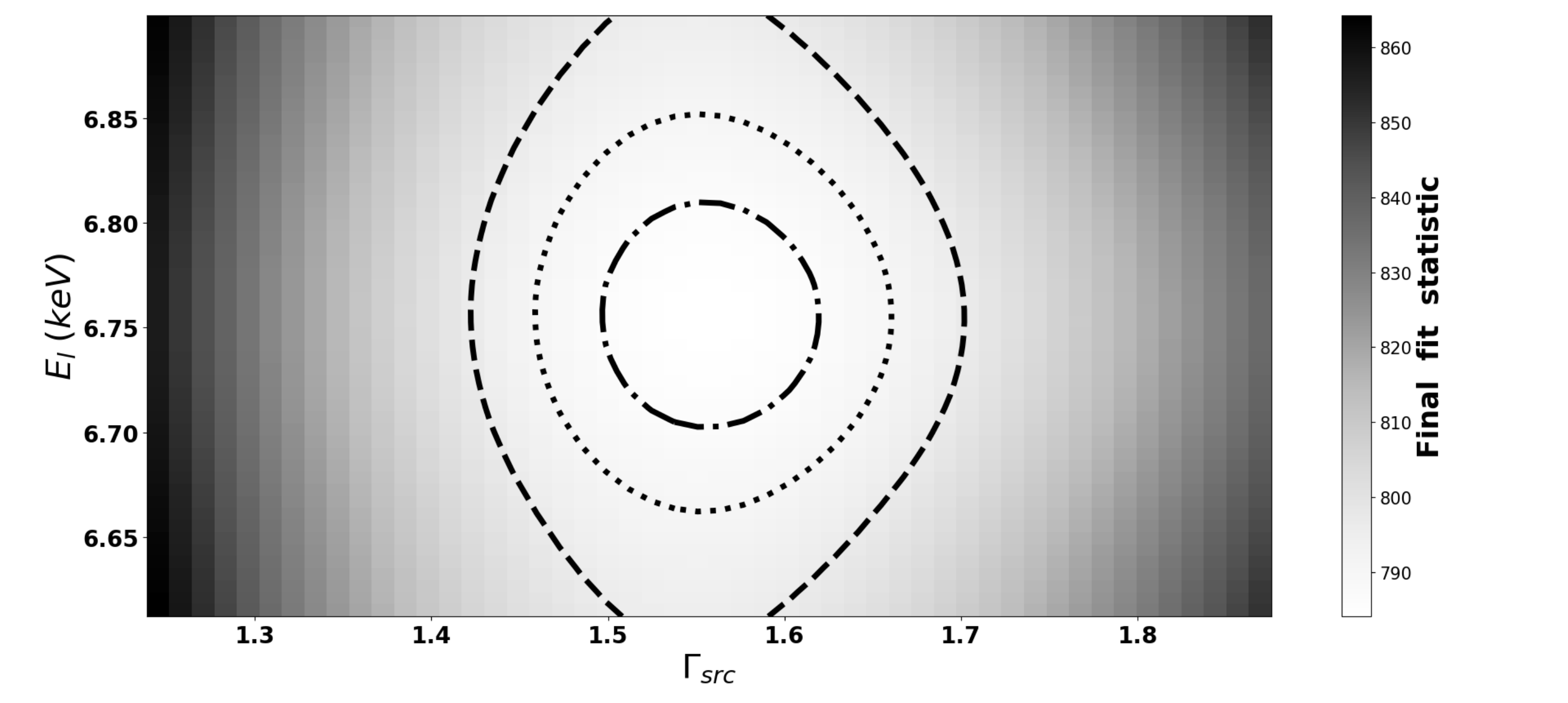}
\hspace{-1.9cm} \includegraphics[width=0.6\textwidth]{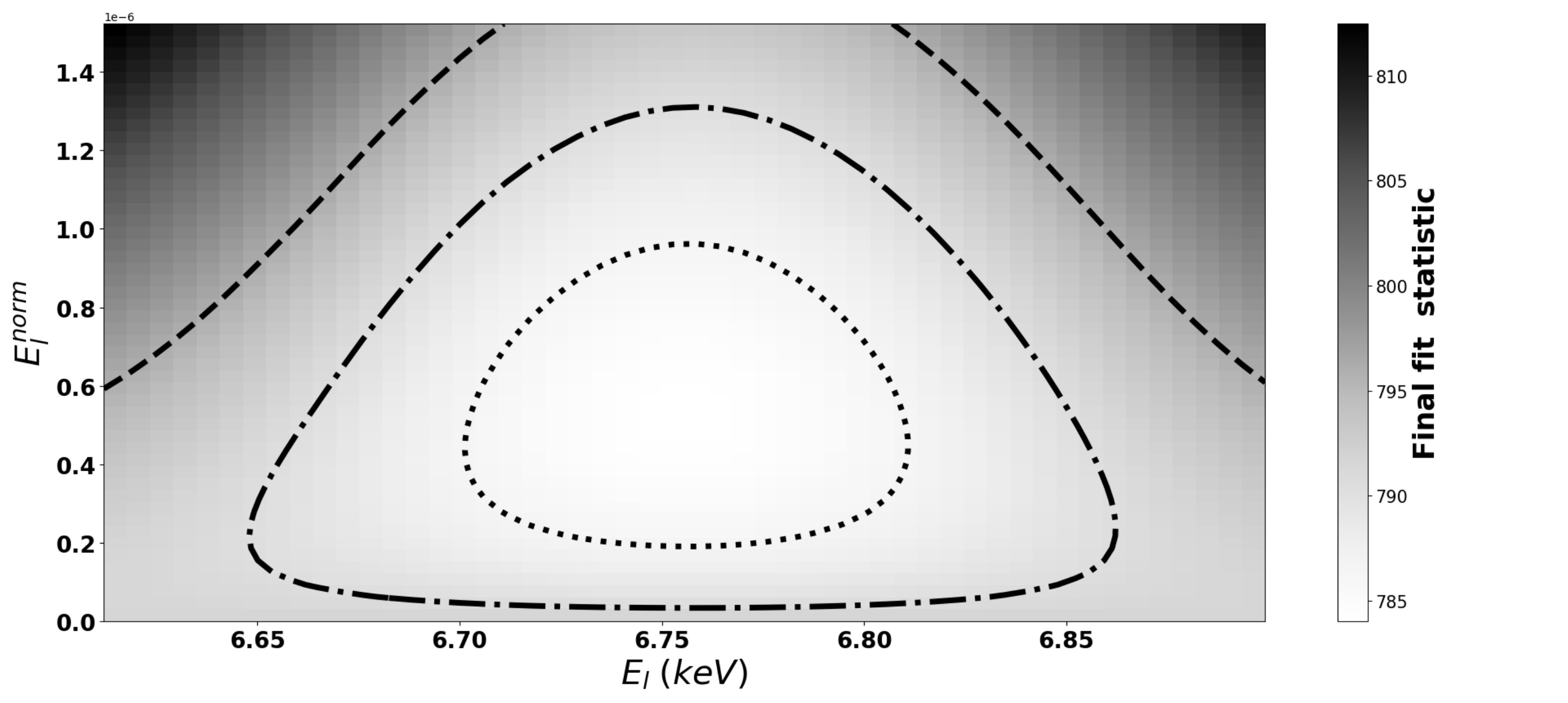}}
\caption{Confidence contours of 1, 2, and 3\,$\sigma$ on the two-thaw-parameters plane for the model F applied to the 0.5--7.0\,keV {\it Chandra} spectrum of CGCG\,292$-$057 nucleus, calculated using the Levenberg-Marquardt optimization method: hydrogen column density $N_{\rm H}^{(1)}$ vs. scattering fraction $f_{sc}$ (upper left panel), power-law photon index $\Gamma$ vs. source-frame line energy $E_l$ (upper right), power-law photon index $\Gamma$ vs. hydrogen column density $N_{\rm H}^{(1)}$ (lower left), source-frame line energy $E_l$ vs. line normalization (lower right).}
\label{fig:cont}
\end{figure*}

The main conclusions emerging from our modeling, as described above, are as follows: (i) simple power-law fits imply very flat, unphysical, photon indices and negligible, although hardly constrained, hydrogen column densities; (ii) a basic thermal model can be rejected, because of the unconstrained, very high, gas temperature; (iii) multi-component models, including two-temperature thermal plasma or thermal plus power-law emission components, do not provide any significant improvement in this respect either; (iv) the source spectrum clearly displays an ionized iron line at $\sim 6.7$\,keV. Keeping this in mind, we introduced our phenomenological ``absorbed scattered power-law'' model (F), which returns very reasonable source parameters and provides a satisfactory fit, and as such, should, in our opinion, be considered as the most plausible option, even though lower and upper $1\,\sigma$ bounds for one of the model parameters could not be constrained with the Nelder-Mead optimization method. In Figure\,\ref{fig:cont} we provide the complementary confidence contour plots of the main model parameters, calculated using the Levenberg-Marquardt optimization method. The unabsorbed $2-10$\,keV flux of the power-law emission component in this model reads as $S_{\rm 2-10\,keV} \simeq (0.73\pm0.20) \times 10^{-13}$\,erg\,cm$^{-2}$\,s$^{-1}$, and the corresponding luminosity as $L_{\rm 2-10\,keV} \simeq 5 \times 10^{41}$\,erg\,s$^{-1}$.

\section{Discussion}

\subsection{Ionized Iron Line}

In type-2 AGN, direct emission of the accretion disk is expected to be observed through a high-column density, cold, circumnuclear gas; some fraction of this emission may, in addition, be scattered by photoionized, optically-thin, medium located at further distances from the center \citep[e.g.,][]{Matt96}. The {\it Chandra} spectrum of the active nucleus in CGCG\,292$-$057 conforms to this general scenario, as it is best fitted with a phenomenological model consisting of a direct, relatively steep-spectrum (photon index $\Gamma \simeq 1.8$), X-ray continuum emission absorbed by a relatively large amount of cold matter (with the equivalent hydrogen column density $N_{\rm H}^{(1)} \simeq 0.7 \times 10^{23}$\,cm$^{-2}$), and partly scattered (fraction $f_{sc} \sim 3\%$) by ionized gas giving rise to a soft excess component (moderated by a relatively small column density of $N_{\rm H}^{(2)} \simeq 10^{21}$\,cm$^{-2}$) and pronounced K$\alpha$ lines from iron ions.

We note that photoionized, Compton-thin gas, scattering the direct X-ray continuum of the accretion disk/disk corona, is not the only viable option for the production of the Fe\,\texttt{XXV} and Fe\,\texttt{XXVI} K$\alpha$ lines in AGN. Alternative models proposed in literature, in this context, include a high-temperature ($>5$\,keV) diffuse thermal plasma \citep{Koyama89,Smith01}, cosmic-ray electrons interacting with a low-temperature ($\sim 0.3$\,keV) thermal plasma \citep{Valinia00,Masai02}, and a Compton-thick accretion disk illuminated by an external X-ray source \citep{Rozanska02,Miller06}. 

Hereafter, however, we exclusively discuss the scenario in which the ionized iron line, detected in the spectrum of CGCG\,292$-$057, results from a Compton-thin gas photoionized by nuclear (direct AGN) illumination, referring, in particular, to the calculations of the resulting line equivalent width performed by \citet{Bianchi02}. In this context, we note that in the Compton-thin regime the ratio between the reflected and transmitted fluxes of the ionizing continuum is expected to be of the order of the Thomson optical depth (assuming the geometrical factor of the order of unity),
\begin{equation}
    \tau \simeq \frac{N_{\rm H}^{(1)}}{\rm 1.5 \times 10^{24}\,cm^{-2}} \sim 0.045 \, ,
\end{equation}
which is indeed very close to the observed (best-fit) value of the scattering fraction $f_{sc} \simeq 0.033$.

The primary parameter in the \citet{Bianchi02} model is the ``X-ray--defined'' ionization parameter, namely
\begin{equation}
    U_x \equiv \frac{\int_{\nu_{\rm L}}^{\nu_{\rm H}} \frac{L_{\nu}}{h\nu} \, d\nu}{4 \pi r^2 c \, n_e} \, ,
    \label{eq:def}
\end{equation}
where $L_{\nu}$ is the luminosity spectral density of the AGN continuum. The integration is over the frequency range corresponding to the photon energy range, $2-10$\,keV, $r$ is the distance of the ionized scattering medium from the source of the photoionizing photons (i.e. from the active nucleus), and $n_e$ is the electron number density characterizing the scattering medium. Since the direct (unabsorbed) AGN continuum is best fitted as a power-law with photon index $\Gamma \simeq 1.8$, the electron number density can be expressed as
\begin{equation}
    n_e \simeq N_{\rm H}^{(1)}/\Delta r \sim 2 \times 10^4 \left(\frac{r}{\rm pc}\right)^{-1} \left(\frac{r}{\Delta r}\right) \, ,
\end{equation}
where $\Delta r$ is the equivalent depth of the ionized absorber involved in re-processing the AGN continuum emission. From equation\,\ref{eq:def} we obtain
\begin{eqnarray}
U_x  & \sim & 30 \, \left(\frac{S_{\rm 2-10\,keV}}{\rm 10^{-13}\,cgs}\right) \left(\frac{r}{\rm pc}\right)^{-2} \left(\frac{n_e}{\rm cm^{-3}}\right)^{-1}  \nonumber \\ & \sim & 10^{-3} \left(\frac{r}{\rm pc}\right)^{-1} \left(\frac{\Delta r}{r}\right) \, .
    \label{eq:Ux}
\end{eqnarray}

Unfortunately, observational constraints on the distance scale $r$ are lacking, with the exception that it must be smaller than the AGN source extraction region in the {\it Chandra} data analysis, namely $r < 5$\,px\,$\simeq 2.5^{\prime\prime} \sim 2.6$\,kpc. On the other hand, a characteristic scale at which one may naturally expect the presence of high-density medium, subjected to a substantial photoionization by the nuclear X-ray emission continuum, is the radius of the ``Broad Line Region'' (BLR), most likely marking a transition between a Compton-thick accretion disk and a clumpy torus of hot circumnuclear dust \citep[see, e.g.,][]{Czerny19}. In order to estimate this radius, in the specific context of CGCG\,292$-$057, first we note that with the given observed flux of the H$\alpha$ emission line of the source \citep{Koziel12,Singh15}, the corresponding luminosity reads as $L_{H\alpha} \simeq 2.5 \times 10^{40}$\,erg\,s$^{-1}$. Hence, the bolometric luminosity of the accreting matter can be found as
\begin{equation}
    L_{\rm bol} \simeq 2 \times 10^3 L_{\rm H\alpha} \sim 5 \times 10^{43}\, {\rm erg\,s^{-1}}
    \label{eq:bol}
\end{equation}
\citep{Netzer09}, and the corresponding Eddington ratio
\begin{equation}
    \lambda_{\rm Edd} \equiv \frac{L_{\rm bol}}{L_{\rm Edd}} \sim (0.7 - 1.4) \times 10^{-3} 
\end{equation}
for the available SMBH mass estimates (see Section\,\ref{sec:intro} above). Given the values for the disk luminosity and accretion rate ---  reinforcing the classification of the analyzed source as a low-luminosity AGN --- we use the best-fit relation between the BLR radius $R_{\rm BLR}$ and the monochromatic disk luminosity $[\lambda L_{\lambda}]_{5100\textup{\AA}}$ as derived by \citet{Bentz13}, namely,
\begin{equation}
    \log \left(\frac{R_{\rm BLR}}{\rm l.d.}\right) = 1.527 + 0.533 \times \log \left(\frac{[\lambda L_{\lambda}]_{5100\textup{\AA}}}{\rm 10^{44}\,erg/s}\right) \, ,
\end{equation} 
where we approximate $[\lambda L_{\lambda}]_{5100\textup{\AA}}$ as simply
\begin{equation}
    [\lambda L_{\lambda}]_{5100\textup{\AA}} \simeq \zeta_{5100\textup{\AA}}^{-1} \, L_{\rm bol} \sim 0.1 \, L_{\rm bol} \, ,
\end{equation}
noting the range for the bolometric correction factor, $\zeta_{5100\textup{\AA}} \simeq 5.5-12.5$, stated in the literature \citep[see][and references therein]{Runnoe12}. With such, the BLR radius in CGCG\,292$-$057 reads as
\begin{equation}
    R_{\rm BLR} \sim 6.8\,{\rm l.d.} \sim 5.7 \times 10^{-3}\,{\rm pc} \, .
\end{equation}

Finally, keeping in mind all the estimates presented above, and assuming $r \sim  R_{\rm BLR}$ as well as, for rather illustrative purposes, $\Delta r \sim r$, we obtain $n_e \sim 3 \times 10^6$\,cm$^{-3}$, and $U_x \sim 0.1$. Given such a value of the ionization parameter, the model calculations by \citet{Bianchi02} --- performed assuming a photon index, of the illuminating continuum, of $2.0$, gas number density $10^6$\,cm$^{-3}$, and gas temperature $10^6$\,K --- for the hydrogen column density of the order of $N_{\rm H}^{(1)}$, imply an Fe\,\texttt{XXV} equivalent width of the order 300\,eV, in excellent agreement with our best-fit value EW\,$\simeq 330^{+210}_{-160}$\,eV. This indicates the scenario involving a Compton-thin gas, located at the distance within the BLR and photoionized by a nuclear illumination, can, in principle, explain the {\it Chandra} observations of CGCG\,292$-$057. If the scattering medium is, however, located at some larger distance from the core, $r \gg  R_{\rm BLR} $, another source of ionization (additional to the AGN illumination) would be needed in order to account for the observed line intensity.

\subsection{Jet---ISM Interactions}

Interestingly, pronounced 6.7\,keV (ionized) iron emission lines, with no accompanying 6.4\,keV (neutral) features, have recently been reported in the spectra of two other restarted/young radio galaxies, namely 0710+43 \citep{Siemiginowska16} and PMN\,J1603--4904 \citep{Muller15,Goldoni16}. Based on those examples (0710+43, PMN\,J1603--4904, and CGCG\,292$-$057), it would be tempting (although premature) to speculate that the ionized iron line emission is somehow related to the presence of compact (galactic-scale) radio lobes of newly triggered relativistic jets. In particular, an expansion of such lobes through the host could lead to the injection of suprathermal/non-thermal electrons into the ISM, providing an additional source of ionization of the iron atoms in the environment \citep[see in this context][]{Masai02}

What is much less speculative, however, is the impact the newly triggered relativistic jets may have on the diffuse phase of the ISM through which the jets propagate. As argued in Section\,\ref{sec:profile}, in the surface brightness profile of CGCG\,292$-$057 we see excess X-ray emission, best fitted as a thermal plasma component with the temperature of $kT_{\rm ISM} \simeq 0.8$\,keV, present at kpc distances from the center. The best-fit normalization of this emission component (see Table\,\ref{tab:ISM}), implies the electron number density,
\begin{equation}
    n_e \simeq \left(\frac{2.2 \times 10^{-6}\, 4 \pi d_{\rm L}^2}{10^{-14} (1+z)^2 \mu V}\right)^{1/2} \sim 5.8 \times 10^{-3} \,{\rm cm^{-3}} ,
\end{equation}
assuming uniform and fully ionized plasma, i.e. $\mu \equiv n_{\rm H}/n_e \simeq 0.82$, within the emission volume
\begin{equation}
    V \simeq \frac{4}{3} \pi \left(R_{\rm out}^3 - R_{\rm in}^3\right) \sim 1,370\,\,{\rm kpc}^3 \, ,
\end{equation}
where $R_{\rm out} = 15$\,px\,$\sim 7.75$\,kpc and $R_{\rm in} = 10$\,px\,$\sim 5.17$\,kpc follows from the applied source extraction region for the ISM emission component. The corresponding gas pressure therefore reads as,
\begin{equation}
    p_{\rm ISM} \simeq n_e \, kT_{\rm ISM} \sim 7.5 \times 10^{-12}\,{\rm erg\,cm^{-3}} \, .
\end{equation}

Since the excess thermal component coincides with the edges of the inner radio structure of the target, it may naturally be explained as resulting from compression and heating of the hot diffuse fraction of the host galaxy, displaced by expanding radio lobes. In such a case, assuming: (i) typical values of the jets' advance velocities $v_{\rm adv} \sim 0.01-0.1 c$, for radio galaxies with linear sizes of the order of $\ell \sim 10$\,kpc \citep[see][]{Kawakatu08}, meaning the jet lifetime $\tau_{\rm j} \sim \ell/v_{\rm adv} \sim 0.3-3$\,Myr, (ii) rough pressure equilibrium between the compact lobes and the surrounding ISM, $p_{\ell}\simeq p_{\rm ISM}$, and (iii) the lobes' volumes are of the same order as $V$ estimated above, then the required jet kinetic power reads as
\begin{equation}
    L_{\rm j}^{\rm (in)} \sim 4 p_{\ell} V / \tau_{\rm j} \sim  10^{43}- 10^{44}\,{\rm erg\,s^{-1}} \, .
\end{equation}
This, in fact, results in a very plausible value, keeping in mind that the bolometric luminosity of the nucleus, as estimated above (equation\,\ref{eq:bol}), implies the mass accretion rate in the source is at least one order of magnitude larger than $L_{\rm j}^{\rm (in)}$, namely
\begin{equation}
    \dot{M}_{\rm acc} c^2 \simeq \eta_{\rm d}^{-1} L_{\rm bol} \sim 10^{45}\,{\rm erg\,cm^{-1}} \,
\end{equation}
where we assumed the radiative disk efficiency $\eta_{\rm d}$ at the level of a few percent, as appropriate at lower accretion rates \citep[see][]{Sharma07}. Moreover, the 1.4\,GHz luminosity spectral density of the outer radio structure in the source, $L_{\rm 1.4\,GHz} \simeq 2 \times 10^{24}$\,W\,Hz$^{-1}$, by means of the calorimetric scaling relation, established for the evolved radio sources by \citet[see also equation 2 in \citealt{Rusinek17}]{Willott99}, implies the comparable jet kinetic power
\begin{equation}
L_{\rm j}^{\rm (out)} = 5.0 \times 10^{22} \, \left(\frac{L_{\rm 1.4\,GHz}}{\rm W\,Hz^{-1}}\right)^{6/7} {\rm erg\,s^{-1}}  
\sim  3 \times 10^{43} {\rm erg\,s^{-1}} \, .
\label{eq:Will}
\end{equation}

The fact that the estimated values of jet kinetic power, supplying the outer and inner structures in this source, are of the same order, is reminiscent of the analogous finding claimed by \citet{Machalski16} for the restarted radio galaxy with MSO-like inner radio structure, 3C\,293.

\subsection{Comparison with Other LINERs}

Narrow K$\alpha$ lines from He- and H-like iron ions are being routinely detected in the X-ray spectra of optically-selected quasars and Seyfert 1s, i.e. radio-quiet type-1 AGN characterized by higher accretion rates \citep{Patrick12}, and have also been reported in the spectra of lower-luminosity AGN such as nearby LINERs, NGC\,4579 and NGC\,4102   \citep{Terashima98,Gonzalez09b,Gonzalez11}. We note that the lines' variability established in NGC\,4579 rules out a thermal (hot ISM plasma) origin of those spectral features \citep[see][]{Terashima00}.

In general, the overall spectral shape of the X-ray radiative output of the CGCG\,292$-$057 nucleus, is reminiscent of the other LINERs studied in X-rays, for which the instruments such as {\it Chandra} or XMM-{\it Newton} reveal power-law emission continua, moderated by intrinsic absorption with a wider range of column densities. Still, the 2--10\,keV luminosity of our target, $\simeq 5 \times 10^{41}$\,erg\,s$^{-1}$, the estimated EW of the Fe\,\texttt{XXV} line, $\simeq 330$\,eV, and the nuclear hydrogen column density $\simeq 0.7 \times 10^{23}$\,cm$^{-2}$, are all at the high ends of the corresponding distributions derived for LINERs \citep[see][]{Gonzalez06,Gonzalez09a,Gonzalez09b}. Also, the BH mass in the system, $\simeq (3-6) \times 10^8 M_{\odot}$, as well as the derived Eddington ratio $ \simeq 10^{-3}$, both seem higher than the average values characterising the LINER population \citep[see][]{Gonzalez09a,Eracleous10}.

Interestingly, the anti-correlation between the X-ray photon index and the 2--10\,keV luminosity (expressed in the Eddington units), as claimed by \citet{Younes11} for LINERs with broad H$\alpha$ emission (equation\,3 therein), in the case of  CGCG\,292$-$057, for which  $L_{\rm 2-10\,keV}/L_{\rm Edd} \simeq 10^{-5}$, yields $\Gamma \simeq 1.66$; this is, in fact, quite close to the best-fit value of $1.8^{+0.42}_{-0.05}$. Yet, at the same time with the bolometric disk luminosity in the system, as given above in equation\,\ref{eq:bol}, we get $L_{\rm bol} / L_{\rm 2-10\,keV} \sim 100$, which is much larger than the bolometric correction factor of the order of 10 claimed by, e.g., \citet{Ho09} or, more recently, \citet{Netzer19}. In other words, the system studied here appears under-luminous in X-rays for its bolometric disk luminosity, when compared with the general population of LINERs. Related, it also appears over-luminous in radio, as its ``X-ray defined radio-loudness parameter'' --- introduced by \citet{Terashima03} in the context of low-luminosity AGN as simply the ratio of the total 5\,GHz luminosity to the 2--10\,keV luminosity --- reads as $\log L_{\rm 5\,GHz}/L_{\rm 2-10\,keV} \simeq -1.2$, meaning a radio-loud object indeed \citep[see also][]{Younes12}. Here we took $L_{\rm 5\,GHz} \simeq 3.8 \times 10^{40}$\,erg\,s$^{-1}$, following from scaling down the 1.4\,GHz luminosity of the outer lobes with the mean radio spectral index of 0.76 \citep{Koziel12}. Note that with such, the ``standard'' radio-loudness parameter, i.e. the ratio of the 5\,GHz and optical $B$-band ($\lambda_B = 4400\textup{\AA}$) luminosity spectral densities,
\begin{equation}
\mathcal{R} \equiv \frac{L_{\nu_5}}{L_{\nu_B}} \simeq 1.4 \times 10^6 \,\, \left(\frac{L_{\rm 5\,GHz}}{L_{\rm bol}}\right) \sim 10^3 \, ,
\end{equation}
places the source on the $\mathcal{R} - \lambda_{\rm Edd}$ plane almost exactly in between the populations of ``Seyferts+LINERs hosted by disk galaxies'' and ``radio galaxies hosted by ellipticals'', as presented and discussed in detail by \citet{Sikora07} in the general context of the AGN jet production efficiency. In the above, we assumed that the monochromatic $B$-band luminosity $[\nu L_{\nu}]_B \simeq 0.1 \times L_{\rm bol}$.

\section{Summary}

The galaxy CGCG\,292$-$057 is a complex, post-merger system at the border between early-type and late-type galaxies, with a LINER-type nucleus and multi-scale/multi-component radio morphology, reflecting intermittent jet activity of the central SMBH. We observed this source with the {\it Chandra} ACIS-S instrument with an exposure totalling 93.8\,ks. 

In our {\it Chandra} exposure, we clearly detected the active nucleus of the system (with the total number of 393 net counts within the $0.5-7.0$\,keV range and 5\,px radius), surrounded by a diffuse, low-surface brightness, emission. We carefully studied the {\it Chandra} PSF and the source radial surface brightness profile, and in this way we found evidence (with significance quantified by the Kolmogorov-Smirnov test) for the presence of an excess X-ray emission component at kpc distances from the center. This component is best fitted by a thermal plasma model with the temperature of $\sim 0.8$\,keV. We argued that the excess emission results from compression and heating of the hot diffuse fraction of the interstellar medium displaced by expanding compact radio jets in the source, for which the kinetic power was estimated as $L_{\rm j} \sim (0.1-1) \times 10^{44}$\,erg\,s$^{-1}$, in agreement with the overall energetics of the studied AGN, and, in particular, the estimated bolometric disk luminosity  $L_{\rm bol} \sim 5 \times 10^{43}$\,erg\,s$^{-1}$ and the mass accretion rate $\dot{M}_{\rm acc} c^2 \sim 10^{45}$\,erg\,s$^{-1}$.

The $0.5-7.0$\,keV spectrum of the CGCG\,292$-$057 nucleus clearly displays an ionized iron line at $\sim 6.7$\,keV (with no accompanying neutral feature at 6.4\,keV). We modeled the spectrum assuming various emission models, and concluded that a simple power-law or either thermal fits imply unphysical photon indices or equivalently very high and unconstrained plasma temperatures. The best fit is obtained assuming a phenomenological model consisting of a relatively steep-spectrum (photon index $\Gamma \simeq 1.8$), direct, X-ray continuum emission absorbed by a relatively large amount of cold matter (with the equivalent hydrogen column density $N_{\rm H}^{(1)} \simeq 0.7 \times 10^{23}$\,cm$^{-2}$), and partly scattered (fraction $f_{sc} \sim 3\%$) by ionized gas, giving rise to a soft excess component and pronounced K$\alpha$ lines from iron He-like ions. The X-ray luminosity of the direct nuclear continuum reads, in this model, as $L_{\rm 2-10\,keV} \simeq 5 \times 10^{41}$\,erg\,s$^{-1}$.

Following detailed calculations by \citet{Bianchi02}, we demonstrated that the scenario involving a Compton-thin gas, located at the distance of the Broad-Line Region in the source ($R_{\rm BLR} \sim 6 \times 10^{-3}$\,pc), and photoionized by nuclear illumination, with the corresponding ionisation parameter $U_x \sim 0.1$, can explain the {\it Chandra} observations of CGCG\,292$-$057, and, in particular, the observed equivalent width of the Fe\,\texttt{XXV} K$\alpha$ line, which is of the order of 0.3\,keV.

We compare the general spectral properties of the CGCG\,292$-$057 nucleus, with those of other nearby LINERs studied in X-rays, and argue that the system appears under-luminous in X-rays for its bolometric disk luministy, and at the same time over-luminous in radio. In fact, on the ``radio-loudness vs. accretion rate'' plane, CGCG\,292$-$057 is located almost exactly in between the populations of Seyferts+LINERs, which are typically hosted by disk galaxies, and radio galaxies, which are typically hosted by ellipticals \citep[see in this context][]{Sikora07}.

\acknowledgments

This work was supported by the Polish NSC grant 2016/22/E/ST9/00061 (K.B., \L .S., V.M., R.T., E.K.) and the Chandra guest investigator program GO5-16109X. This research was supported in part by NASA through contract NAS8-03060 (A.S., M.S.) to the Chandra X-ray Center. Work by C.C.C. at NRL is supported in part by NASA DPR S-15633-Y.\\ K.B. and \L .S. thank Marek Jamrozy, Dorota Kozie{\l}-Wierzbowska, and Swayamtrupta Panda for useful suggestions and discussions on the manuscripts. The authors also acknowledge the anonymous referee for her/his careful reading of the manuscript and useful suggestions, which helped to improve the presentation.
\vspace{5mm}
\facilities{Chandra (ACIS)}
\software{CIAO \citep{Fruscione06}, Sherpa \citep{Freeman01}, ChaRT \citep{Carter03}, and MARX \citep{Davis12}}

\appendix

\section{Modeling the Chandra PSF}
\label{appendix}

We performed High Resolution Mirror Assembly (HRMA) PSF simulations for the core, using the Chandra Ray Tracer (ChaRT) online tool \citep{Carter03}\footnote{\url{http://cxc.harvard.edu/ciao/PSFs/chart2/runchart.html}}. For these simulations, the source spectrum file (corresponding to the model fit A, i.e. absorbed power-law; see Table\,\ref{tab:AGN}) was uploaded to ChaRT, obtaining a set of rays, which were next projected onto the detector plane with the MARX software \citep{Davis12}\footnote{\url{https://space.mit.edu/cxc/marx}}. The resulting PSF files were normalized to the observed count rate, and filtered with the source region at a bin factor of 1. 

\begin{figure*}[!t]
\centering
\includegraphics[width=0.49\textwidth]{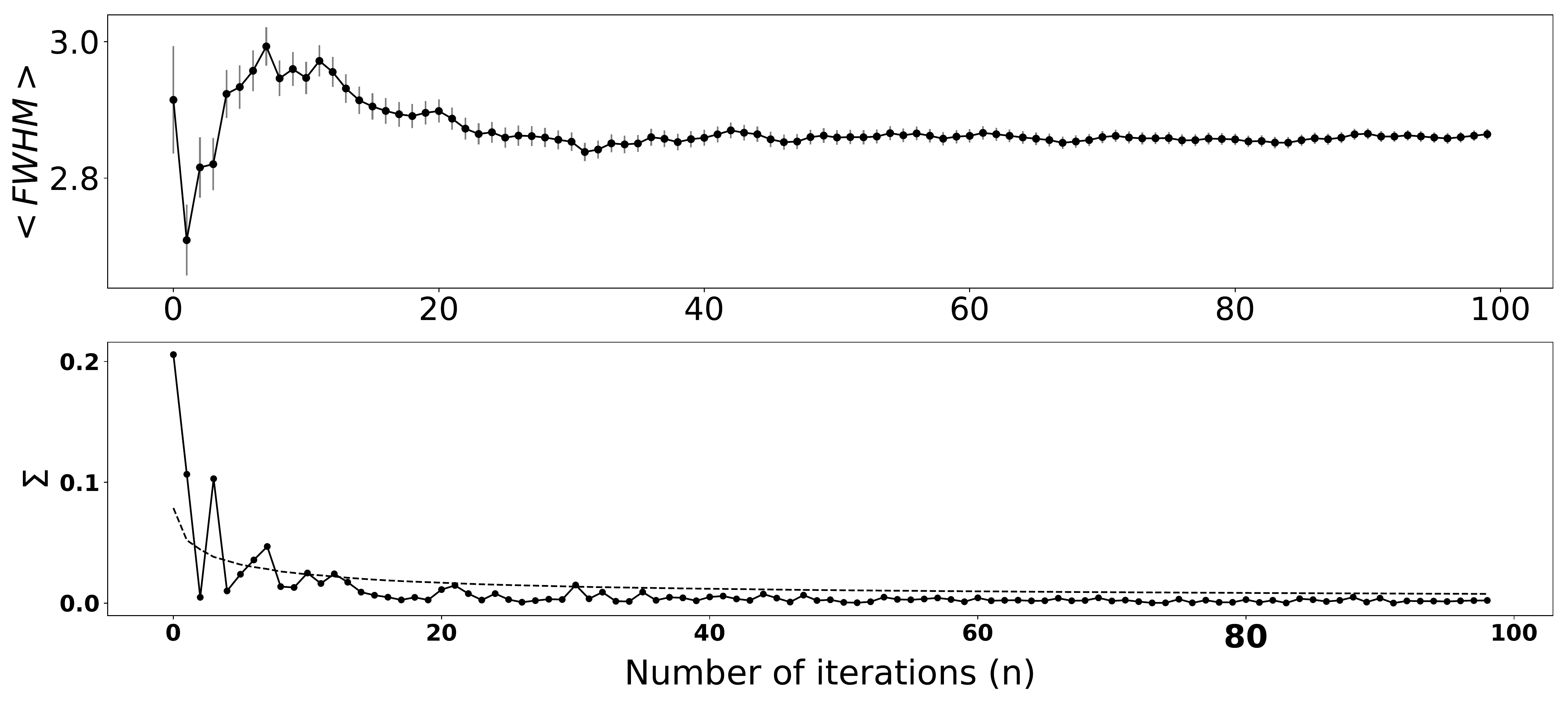}
\includegraphics[width=0.49\textwidth]{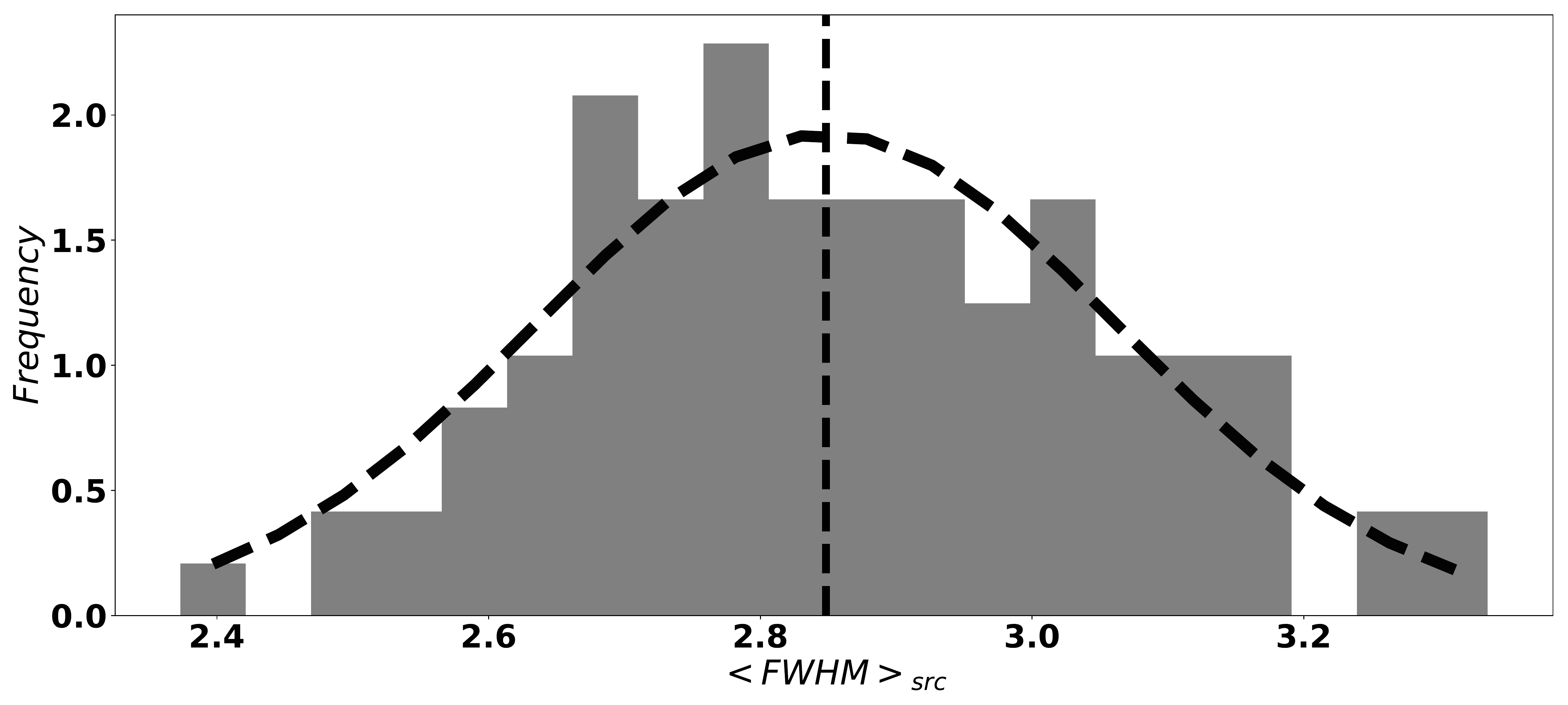}
\caption{{\it Left:} the estimated mean FWHM of the PSF modeled with 2D Gaussian, $\langle {\rm FWHM} \rangle_{\rm N} \pm \Sigma_{\rm N}$, as a function of the number N of PSF simulations performed (upper panel); the successive difference in the mean values $|\langle {\rm FWHM} \rangle_{\rm N}-\langle {\rm FWHM} \rangle_{\rm{N-1}}|$, along with the standard error of the mean $\Sigma_{\rm N}$ (dashed curve), are shown on the lower panel.\\
{\it Right:} Histogram of the emerging FWHM values after 100 PSF simulations, along with the best-fit normal distribution returning $\langle {\rm FWHM} \rangle_{100} \simeq 2.84$\,px and $\Sigma \simeq 0.21$\,px.}
\label{fig:iterations}
\end{figure*}

Because of limited photon statistics, it is reasonable to expect quite substantial differences in each particular realization of the PSF due to random photon fluctuations. To investigate this effect in more detail, we repeated PSF simulations several times, and modeled the resulting PSF profiles with a 2D Gaussian. In this way, we estimate the Full Width at Half Maximum (FWHM) of the central point source in the given image. We then investigated the point of convergence for the fluctuations in the PSF simulations by deriving the mean values, $\langle {\rm FWHM} \rangle_{\rm N} \pm \Sigma_{\rm N}$, for an increasing number, N, of PSF simulations performed, as well as the successive difference in the mean values of FWHM, namely $|\langle {\rm FWHM} \rangle_{\rm N}-\langle {\rm FWHM} \rangle_{\rm{N-1}}|$. The results of this analysis, for N up to 100, are presented in the left panel of Figure\,\ref{fig:iterations}. As shown in the figure, the mean FWHM converges to $\simeq 2.85$\,px after only about 20 iterations, N, at which point the standard error of the mean, which decreases as $\Sigma_{\rm N} \simeq \Sigma/\sqrt{\rm N}$ (as in fact expected for a normal distribution of FWHM values with a standard deviation of the population $\Sigma$) exceeds the differential error $|\langle {\rm FWHM} \rangle_{\rm N}-\langle {\rm FWHM} \rangle_{\rm{N-1}}|$. 

\begin{figure*}[!h]
\centering
\includegraphics[width=\textwidth]{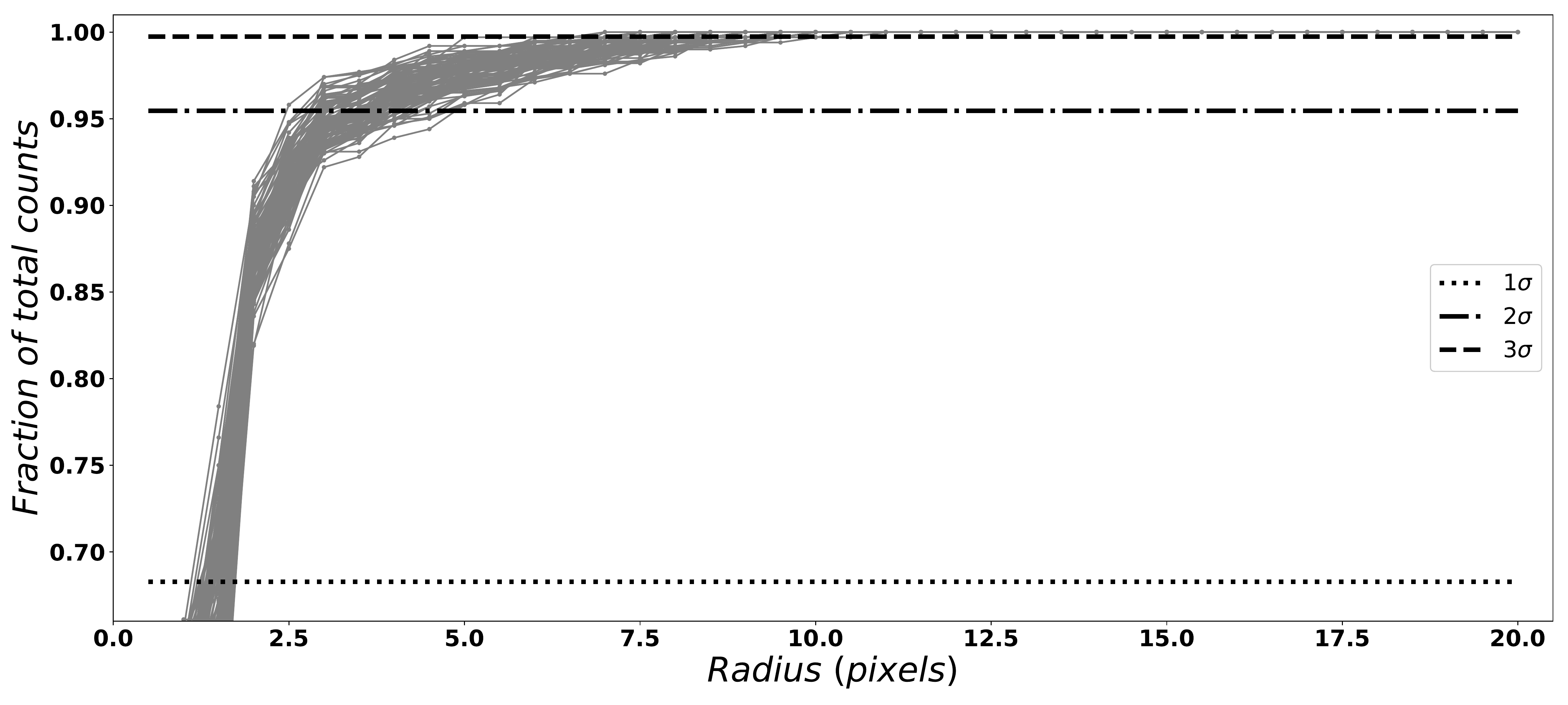}
\caption{Enclosed count fraction for 100 simulated PSF images; horizontal dotted lines from bottom to top correspond to $1\sigma$, $2\sigma$, and $3\sigma$ count fractions, respectively.}
\label{fig:PSF}
\end{figure*}

Right panel of Figure\,\ref{fig:iterations} presents the histogram of the emerging FWHM values after 100 PSF simulations along with the best-fit normal distribution returning, $\langle {\rm FWHM} \rangle_{100} \simeq 2.84$\,px and $\Sigma \simeq 0.21$\,px. Given this, the radius for the extraction region encompassing $3\sigma$ of the photons from a point source formally reads as, $\simeq (3/2 \sqrt{2 \ln 2}) \times \frac{1}{2} \langle {\rm FWHM} \rangle_{100} \simeq 1.8\pm0.1$\,px. However, the source PSF is not well represented by a 2D Gaussian and particularly with limited photon statistics, broad wings of the PSF do matter. To illustrate this, in Figure\,\ref{fig:PSF} we plot the enclosed photon count fraction for each simulated PSF. As shown, the real $2\sigma$ radius ranges between 2.5\,px and 5\,px, depending on a particular PSF realization, and the $3\sigma$ radius is, in all cases, at $\sim 10$\,px. Hence, in the analysis of the surface brightness presented in this paper, instead of a 2D Gaussian approximation, we build a table model for the core PSF as a result from the averaging of over 100 PSF simulations.

\section{Surface Brightness Profiles Along and Across the Jet Axis}
\label{appendixB}

\begin{figure}[!t]
\centering
\includegraphics[width=0.49\textwidth]{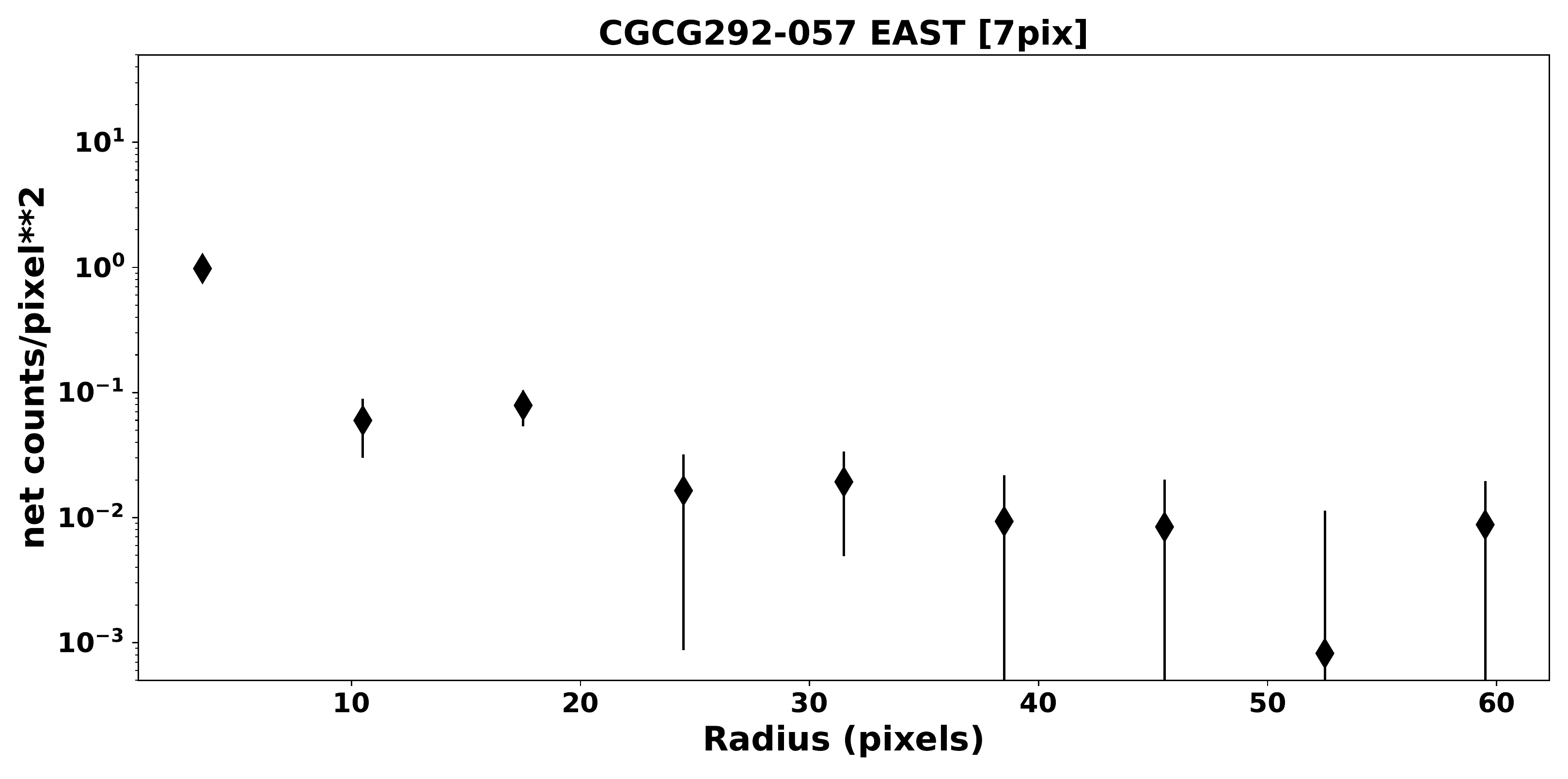}
\includegraphics[width=0.49\textwidth]{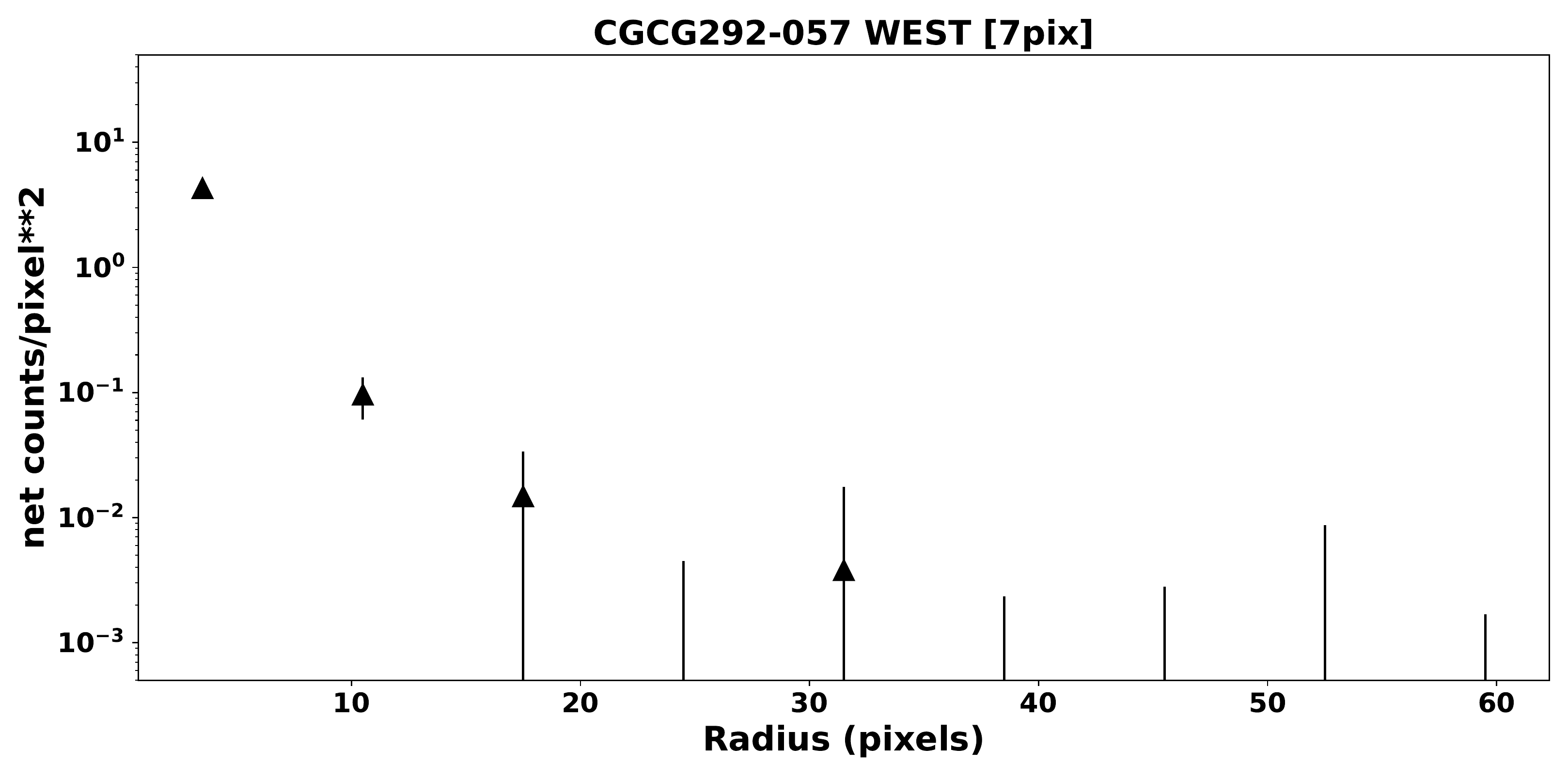}
\includegraphics[width=0.49\textwidth]{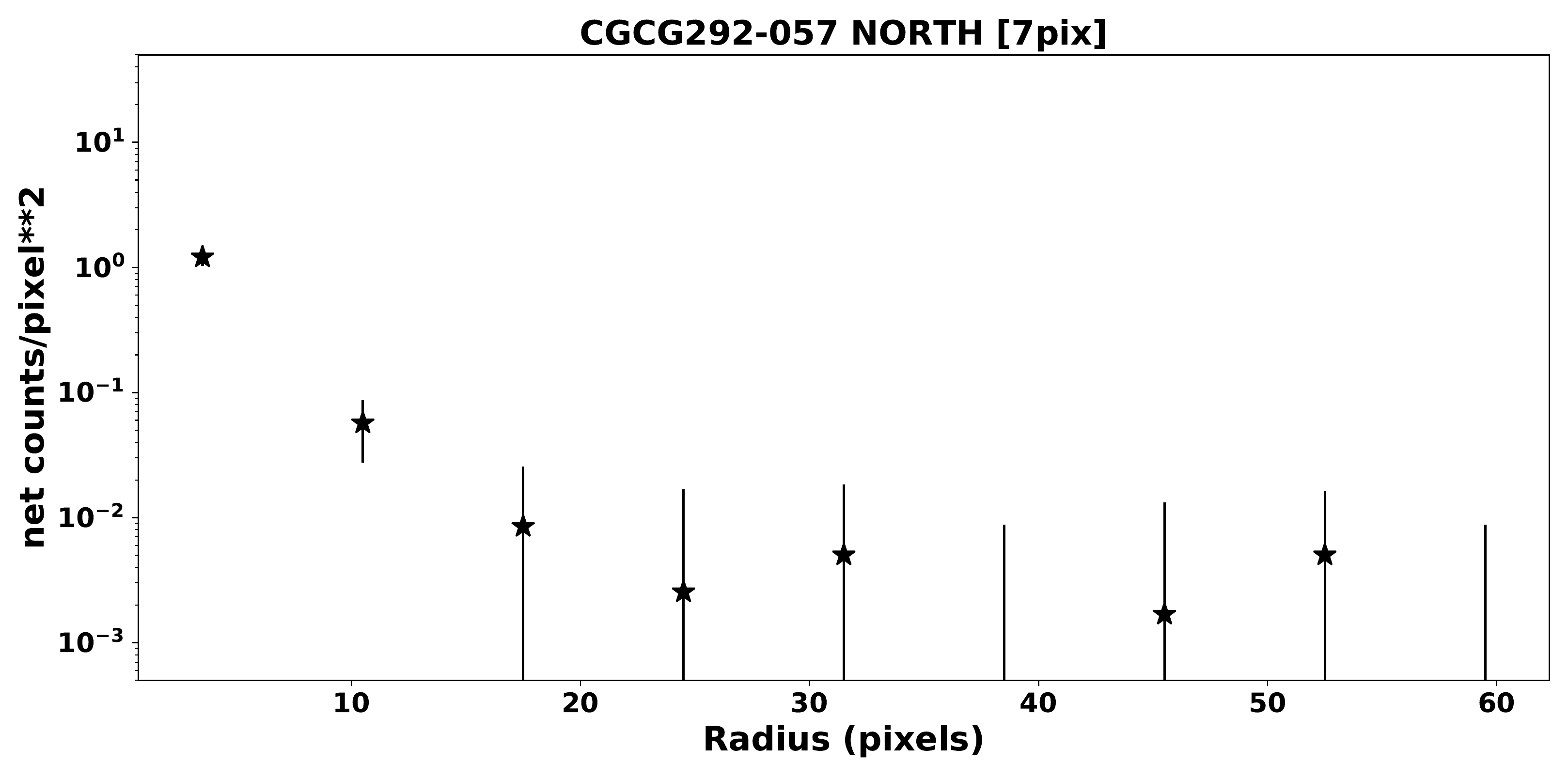}
\includegraphics[width=0.49\textwidth]{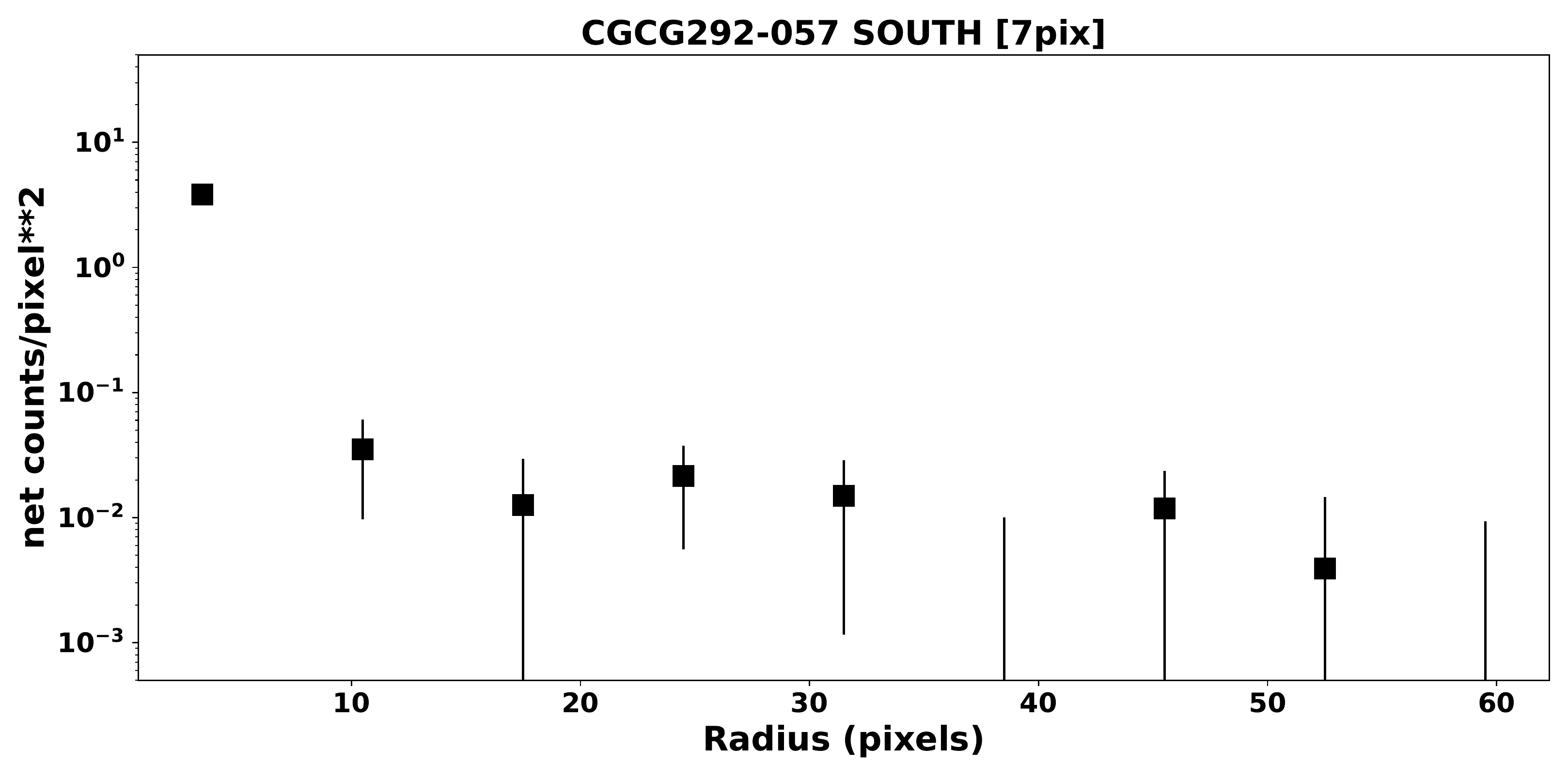}
\caption{X-ray surface brightness profiles resulting from extracting the net counts from four quadrants of a concentric stack of annular regions (each with the bin size of 7\,px) centered on the CGCG\,292$-$057 nucleus. The quadrants are oriented to the East, to the West, to the North, and to the South from the center (top-left, top-right, bottom-left, and bottom-right panels, respectively).}
\label{fig:NEWS}
\end{figure}

We have also extracted the radial profiles of the net counts separately for the four quadrants of the concentric annular regions centered on the CGCG\,292$-$057 nucleus, oriented to the East, to the West, to the North, and to the South from the center. The East and West quadrants correspond therefore to the surface brightness profiles \emph{along} the axis of the inner jets in the system, while the North and South quadrants to the surface brightness profiles \emph{across} the jet axis. Due to the limited photon statistics in separate quadrants, we had to increase the size of the annuli from 2\,px (as adopted in Figure\,\ref{fig:SB}) up to 7\,px. The resulting profiles are presented in Figure\,\ref{fig:NEWS}.

\begin{thebibliography}{}

\bibitem[Ahn et al.(2014)]{Ahn14} Ahn, C.~P., Alexandroff, R., Allende Prieto, C., et al.\ 2014, \apjs, 211, 17

\bibitem[Augusto et al.(2006)]{Augusto06} Augusto, P., Gonzalez-Serrano, J.~I., Perez-Fournon, I., et al.\ 2006, \mnras, 368, 1411

\bibitem[Balasubramaniam et al.(2017)]{Balasubramaniam17} Balasubramaniam, K., Stawarz, L., Sobolewska, M., et al.\ 2017, The X-ray Universe 2017, 251

\bibitem[Bentz et al.(2013)]{Bentz13} Bentz, M.~C., Denney, K.~D., Grier, C.~J., et al.\ 2013, \apj, 767, 149

\bibitem[Bentz et al.(2009)]{Bentz09} Bentz, M.~C., Peterson, B.~M., Pogge, R.~W., et al.\ 2009, \apjl, 694, L166

\bibitem[Beuchert et al.(2018)]{Beuchert18} Beuchert, T., Rodr{\'\i}guez-Ardila, A., Moss, V.~A., et al.\ 2018, \aap, 612, L4

\bibitem[Bianchi \& Matt(2002)]{Bianchi02} Bianchi, S., \& Matt, G.\ 2002, \aap, 387, 76

\bibitem[Carter et al.(2003)]{Carter03} Carter, C., Karovska, M., Jerius, D., et al.\ 2003, Astronomical Data Analysis Software and Systems XII, 477

\bibitem[Cheung et al.(2009)]{Cheung09} Cheung, C.~C., Healey, S.~E., Landt, H., et al.\ 2009, \apjs, 181, 548

\bibitem[Cheung(2007)]{Cheung07} Cheung, C.~C.\ 2007, \aj, 133, 2097

\bibitem[Czerny(2019)]{Czerny19} Czerny, B.\ 2019, Open Astronomy, 28, 200

\bibitem[Czerny et al.(2009)]{Czerny09} Czerny, B., Siemiginowska, A., Janiuk, A., et al.\ 2009, \apj, 698, 840

\bibitem[Davis et al.(2012)]{Davis12} Davis, J.~E., Bautz, M.~W., Dewey, D., et al.\ 2012, \procspie, 84431A

\bibitem[Eracleous et al.(2010)]{Eracleous10} Eracleous, M., Hwang, J.~A., \& Flohic, H.~M.~L.~G.\ 2010, \apjs, 187, 135

\bibitem[Freeman et al.(2001)]{Freeman01} Freeman, P., Doe, S., \& Siemiginowska, A.\ 2001, \procspie, 4477, 76

\bibitem[Fruscione et al.(2006)]{Fruscione06} Fruscione, A., McDowell, J.~C., Allen, G.~E., et al.\ 2006, \procspie, 6270, 62701V

\bibitem[Goldoni et al.(2016)]{Goldoni16} Goldoni, P., Pita, S., Boisson, C., et al.\ 2016, \aap, 586, L2

\bibitem[Gonz{\'a}lez-Mart{\'\i}n et al.(2011)]{Gonzalez11} Gonz{\'a}lez-Mart{\'\i}n, O., Papadakis, I., Braito, V., et al.\ 2011, \aap, 527, A142

\bibitem[Gonz{\'a}lez-Mart{\'\i}n et al.(2009b)]{Gonzalez09b} Gonz{\'a}lez-Mart{\'\i}n, O., Masegosa, J., M{\'a}rquez, I., et al.\ 2009b, \aap, 506, 1107

\bibitem[Gonz{\'a}lez-Mart{\'\i}n et al.(2009a)]{Gonzalez09a} Gonz{\'a}lez-Mart{\'\i}n, O., Masegosa, J., M{\'a}rquez, I., et al.\ 2009a, \apj, 704, 1570

\bibitem[Gonz{\'a}lez-Mart{\'\i}n et al.(2006)]{Gonzalez06} Gonz{\'a}lez-Mart{\'\i}n, O., Masegosa, J., M{\'a}rquez, I., et al.\ 2006, \aap, 460, 45

\bibitem[Ho(2009)]{Ho09} Ho, L.~C.\ 2009, \apj, 699, 626

\bibitem[Ho(2008)]{Ho08} Ho, L.~C.\ 2008, \araa, 46, 475

\bibitem[Hopkins et al.(2006)]{Hopkins06} Hopkins, P.~F., Hernquist, L., Cox, T.~J., et al.\ 2006, \apjs, 163, 1

\bibitem[Kawakatu et al.(2008)]{Kawakatu08} Kawakatu, N., Nagai, H., \& Kino, M.\ 2008, \apj, 687, 141

\bibitem[Kennicutt(1998)]{Kennicutt98} Kennicutt, R.~C.\ 1998, \araa, 36, 189

\bibitem[Konar et al.(2019)]{Konar19} Konar, C., Hardcastle, M.~J., Croston, J.~H., et al.\ 2019, \mnras, 486, 3975

\bibitem[Koyama et al.(1989)]{Koyama89} Koyama, K., Awaki, H., Kunieda, H., et al.\ 1989, \nat, 339, 603

\bibitem[Kozie{\l}-Wierzbowska et al.(2012)]{Koziel12} Kozie{\l}-Wierzbowska, D., Jamrozy, M., Zola, S., et al.\ 2012, \mnras, 422, 1546

\bibitem[Machalski et al.(2016)]{Machalski16} Machalski, J., Jamrozy, M., Stawarz, {\L}., et al.\ 2016, \aap, 595, A46

\bibitem[Mahatma et al.(2019)]{Mahatma19} Mahatma, V.~H., Hardcastle, M.~J., Williams, W.~L., et al.\ 2019, \aap, 622, A13

\bibitem[Masai et al.(2002)]{Masai02} Masai, K., Dogiel, V.~A., Inoue, H., et al.\ 2002, \apj, 581, 1071

\bibitem[Matt et al.(1996)]{Matt96} Matt, G., Brandt, W.~N., \& Fabian, A.~C.\ 1996, \mnras, 280, 823

\bibitem[McConnell \& Ma(2013)]{McConnell13} McConnell, N.~J. \& Ma, C.-P.\ 2013, \apj, 764, 184

\bibitem[Miller et al.(2006)]{Miller06} Miller, L., Turner, T.~J., Reeves, J.~N., et al.\ 2006, \aap, 453, L13

\bibitem[M{\"u}ller et al.(2015)]{Muller15} M{\"u}ller, C., Krau{\ss}, F., Dauser, T., et al.\ 2015, \aap, 574, A117

\bibitem[Netzer(2019)]{Netzer19} Netzer, H.\ 2019, \mnras, 488, 5185

\bibitem[Netzer(2009)]{Netzer09} Netzer, H.\ 2009, \mnras, 399, 1907

\bibitem[O'Dea, \& Baum(1997)]{odea97} O'Dea, C.~P., \& Baum, S.~A.\ 1997, \aj, 113, 148

\bibitem[Patrick et al.(2012)]{Patrick12} Patrick, A.~R., Reeves, J.~N., Porquet, D., et al.\ 2012, \mnras, 426, 2522

\bibitem[Reynolds \& Begelman(1997)]{Reynolds97} Reynolds, C.~S., \& Begelman, M.~C.\ 1997, \apjl, 487, L135

\bibitem[Reynolds et al.(2001)]{Reynolds01} Reynolds, C.~S., Heinz, S., \& Begelman, M.~C.\ 2001, \apjl, 549, L179

\bibitem[R{\'o}{\.z}a{\'n}ska et al.(2002)]{Rozanska02} R{\'o}{\.z}a{\'n}ska, A., Dumont, A.-M., Czerny, B., et al.\ 2002, \mnras, 332, 799

\bibitem[Runnoe et al.(2012)]{Runnoe12} Runnoe, J.~C., Brotherton, M.~S., \& Shang, Z.\ 2012, \mnras, 422, 478

\bibitem[Rusinek et al.(2017)]{Rusinek17} Rusinek, K., Sikora, M., Kozie{\l}-Wierzbowska, D., \& Godfrey, L.\ 2017, \mnras, 466, 2294 

\bibitem[Sharma et al.(2007)]{Sharma07} Sharma, P., Quataert, E., Hammett, G.~W., et al.\ 2007, \apj, 667, 714

\bibitem[Shimasaku et al.(2001)]{Shimasaku01} Shimasaku, K., Fukugita, M., Doi, M., et al.\ 2001, \aj, 122, 1238

\bibitem[Siemiginowska et al.(2016)]{Siemiginowska16} Siemiginowska, A., Sobolewska, M., Migliori, G., et al.\ 2016, \apj, 823, 57

\bibitem[Sikora et al.(2007)]{Sikora07} Sikora, M., Stawarz, {\L}., \& Lasota, J.-P.\ 2007, \apj, 658, 815

\bibitem[Singh et al.(2015)]{Singh15} Singh, V., Ishwara-Chandra, C.~H., Sievers, J., et al.\ 2015, \mnras, 454, 1556

\bibitem[Smith et al.(2001)]{Smith01} Smith, R.~K., Brickhouse, N.~S., Liedahl, D.~A., et al.\ 2001, \apjl, 556, L91

\bibitem[Stawarz et al.(2008)]{Stawarz08} Stawarz, {\L}., Ostorero, L., Begelman, M.~C., et al.\ 2008, \apj, 680, 911

\bibitem[Stoughton et al.(2002)]{Stoughton02} Stoughton, C., Lupton, R.~H., Bernardi, M., et al.\ 2002, \aj, 123, 485

\bibitem[Strateva et al.(2001)]{Strateva01} Strateva, I., Ivezi{\'c}, {\v{Z}}., Knapp, G.~R., et al.\ 2001, \aj, 122, 1861

\bibitem[Terashima et al.(2000)]{Terashima00} Terashima, Y., Ho, L.~C., Ptak, A.~F., et al.\ 2000, \apjl, 535, L79

\bibitem[Terashima et al.(1998)]{Terashima98} Terashima, Y., Kunieda, H., Misaki, K., et al.\ 1998, \apj, 503, 212

\bibitem[Terashima \& Wilson(2003)]{Terashima03} Terashima, Y., \& Wilson, A.~S.\ 2003, \apj, 583, 145

\bibitem[Tremaine et al.(2002)]{Tremaine02} Tremaine, S., Gebhardt, K., Bender, R., et al.\ 2002, \apj, 574, 740

\bibitem[Valinia et al.(2000)]{Valinia00} Valinia, A., Tatischeff, V., Arnaud, K., et al.\ 2000, \apj, 543, 733

\bibitem[Willott et al.(1999)]{Willott99} Willott, C.~J., Rawlings, S., Blundell, K.~M., \& Lacy, M.\ 1999, \mnras, 309, 1017 

\bibitem[Younes et al.(2012)]{Younes12} Younes, G., Porquet, D., Sabra, B., et al.\ 2012, \aap, 539, A104

\bibitem[Younes et al.(2011)]{Younes11} Younes, G., Porquet, D., Sabra, B., et al.\ 2011, \aap, 530, A149

\end{thebibliography}
\end{document}